\documentclass[%
 reprint,
amsmath,amssymb,
aps,
]{revtex4-2}

\usepackage{graphicx}% Include figure files
\usepackage{dcolumn}% Align table columns on decimal point
\usepackage{bm}% bold math
\usepackage{orcidlink} %To show orchid link
\usepackage{float}

\begin{document}

\title{\boldmath Probing mass orderings in presence of a very light sterile neutrino in a liquid argon detector}

% more complex case: 4 authors, 3 institutions, 2 footnotes
\author{Animesh~Chatterjee$^{1,2}$ \orcidlink{0000-0002-2935-0958}}
\email{animesh.chatterjee@cern.ch}
\author{Srubabati~Goswami$^1$ \orcidlink{0000-0002-5614-4092}}
\email{sruba@prl.res.in}
\author{Supriya~Pan$^{1,3}$ \orcidlink{0000-0003-3556-8619}}
\email{supriyapan@prl.res.in}
% The "\note" macro will give a warning: "Ignoring empty anchor..."
% you can safely ignore it.

\affiliation{$^1$ Physical Research Laboratory, Ahmedabad, Gujarat, 380009, India}
\affiliation{$^2$ European Organization for Nuclear Research (CERN), 1211 Geneva 23, Switzerland}
\affiliation{$^3$ Indian Institute of Technology, Gandhinagar, Gujarat, 382355, India}

\begin{abstract}
    Results from experiments like LSND and MiniBooNE hint towards the possible presence of an extra eV scale sterile neutrino. The addition of such a neutrino will significantly impact the standard three flavour neutrino oscillations. In particular, it can give rise to additional degeneracies due to additional sterile parameters. For an eV scale sterile neutrino, the cosmological constraints dictate that the sterile state is heavier than the three active states. However, for lower masses of sterile neutrinos, it can be lighter than one and/or more of the three states. In such cases, the mass ordering of the sterile neutrinos also becomes unknown along with the mass ordering of the active states. In this paper, we explore the mass ordering sensitivity in the presence of a sterile neutrino assuming the mass squared difference $|\Delta_{41}|$ to be in the range $10^{-4} - 1$ eV$^2$. We study (i) how the ordering of the active states, i.e. the determination of the sign of $\Delta_{31}$ gets affected by the presence of a sterile neutrino in the above mass range, (ii) the possible determination of the sign of $\Delta_{41}$ for $\Delta_{41}$ in the range $10^{-4} - 0.1$ eV$^2$. This analysis is done in the context of a liquid argon detector using both beam neutrinos traveling a distance of 1300 km and atmospheric neutrinos which propagates through a distance ranging from 10 - 10000 km allowing resonant matter effects. Apart from presenting  separate results from these sources, we also do a combined study and probe the synergy between these two in giving an enhanced sensitivity.
\end{abstract}

\maketitle
\flushbottom

\section{Introduction}
\label{sec:intro}
    Neutrino oscillations in which one flavour of neutrinos gets converted to others, have been discovered using diverse neutrino sources: from the sun to the atmosphere, from reactors to accelerators, and a variety of detection techniques in different terrestrial experiments. This has validated the three neutrino oscillation picture and most of the parameters governing the phenomenon have been determined with sizable precision. These parameters are the mass squared differences $\Delta_{21}$, $|\Delta_{31}|$ ($\Delta_{ij}$ is defined as $m_i^2-m_j^2$ for the mass eigenvalues $m_i,m_j$), and mixing angles $\theta_{12}$, $\theta_{23}$. The remaining unknowns of the three flavour paradigm are the atmospheric mass ordering (the sign of $\Delta_{31}$), the octant of the mixing angle $\theta_{23}$, and the CP phase $\delta_{CP}$. The next generation beam based experiments like DUNE \cite{DUNE:2016ymp, DUNE:2020ypp}, T2HK \cite{Hyper-Kamiokande:2016dsw}, ESS$\nu$SB \cite{Alekou:2022emd} are expected to resolve these issues. These experiments plan to use a beam of higher intensity and larger volume detectors which can increase the statistics as compared to the current generation accelerator experiments T2K \cite{T2K:2019bcf} and NO$\nu$A\cite{NOvA:2007rmc}. Apart from these there are dedicated future atmospheric neutrino experiments like Hyper Kamiokande\cite{Hyper-KamiokandeProto-:2015xww}, India-based Neutrino Observatory \cite{ICAL:2015stm}, IceCube and Pingu\cite{ref:IceCube} which can also help in throwing light on the final frontier of the three neutrino oscillation. The liquid argon detector in DUNE will also be able to observe atmospheric neutrinos \cite{Barger:2013rha,Kelly:2021jfs,Arguelles:2022hrt}. Apart from the determination of parameters in the standard three flavour oscillation, these next generation experiments also open up the possibility of probing beyond standard model (BSM) physics, which can occur at a sub-leading level. Several new physics scenarios like light sterile neutrinos, non-standard interactions, long-range forces, neutrino decay, and violation of fundamental symmetries like CPT, Lorentz-invariance, etc have been explored in the context of neutrino oscillation experiments\cite{DUNE:2020fgq,Arguelles:2019xgp}. 

    Among these, the light sterile neutrino scenario is motivated by three long-standing anomalies that have served as primary drivers in the development of a vibrant short-baseline neutrino program over the last decade. Two of these pieces of evidence come from the apparent oscillatory appearance of electron (anti)neutrinos in muon-(anti)neutrino beams originating from charged-pion decay-at-rest in the LSND experiment \cite{LSND:2001aii, KARMEN:2002zcm}, and charged-pion decay-in-flight in the MiniBooNE experiment \cite{MiniBooNE:2008paa, MiniBooNE:2007uho}. There have also been an anomaly associated with an overall normalization discrepancy of electron (anti)neutrinos expected in the radioactive decay of Gallium-71\cite{GALLEX:1997lja, SAGE:1998fvr, Barinov:2022wfh}. One of the most theoretically motivated frameworks considered for the interpretation of these anomalies is that of the presence of light ($\approx$ 1 eV) sterile neutrino state\cite{Abazajian:2012ys}.

    A sterile neutrino is a neutral SU(2)$\times$U(1) singlet with no ordinary weak interaction except those induced by the mixing. Very heavy sterile neutrinos ($10^{14} - 10^{16}$ GeV) are proposed as the mediators in the type I seesaw model\cite{Minkowski:1977sc, Mohapatra:1979ia, Schechter:1980gr} which can give rise to small neutrino masses. Such neutrinos also play a significant role in leptogenesis\cite{Fukugita:1986hr,Davidson:2008bu}. Such neutrinos are natural candidates in grand unified theories. Sterile neutrinos of TeV energies have also been studied in the context of low-scale seesaw models\cite{Appelquist:2002me, Appelquist:2003uu}. Sterile neutrinos of keV mass are especially interesting because the sterile neutrinos would be a viable dark matter candidate\cite{Dodelson:1993je}.

    The existence of an eV scale sterile neutrino  motivated by the short-baseline anomalies is in strong tension from cosmological bound on the effective number of neutrinos $N_{eff}$ and the sum of total masses of light neutrinos. From the recent measurement of Planck data\cite{Planck:2018vyg} and combining together with the Hubble parameter measurement \cite{Riess_2018} and Supernova Ia data from the Pantheon sample \cite{Pan-STARRS1:2017jku}, the extended fit to the parameters are
    \begin{equation}
    \begin{split}
        N_{eff} & = 3.11^{+0.37}_{-0.36}   (95\%  \text{ CL})\\
       & \sum m_{\nu} <   0.16 \text{ eV}
    \end{split}
    \end{equation}
    In order to get around cosmological constraints, we need to introduce new physics that directly affects the cosmological phenomenology of the light sterile states. Since the main problem of the canonical light sterile neutrino is that its thermalization in the early universe raises $N_{eff}$ to an unacceptably large level for BBN and CMB/LSS constraints.  All known new physics solutions so far involve tampering with the thermalization process, in order to maintain $N_{eff}$ as close to the SM value as possible. A number of ideas have been proposed and explored throughout the years, e.g., large chemical potentials or, equivalently, number density asymmetries for the active neutrinos\cite{Foot:1995bm}; secret interactions  of the sterile neutrinos\cite{Dasgupta:2013zpn, Chu:2018gxk} and low reheating temperature of the universe\cite{Yaguna:2007wi}, etc.
    
    Can there be sterile neutrinos lighter than the eV scale? In the presence of a sterile neutrino, there is a new mass squared difference $\Delta_{41}=m_4^2-m_1^2$. A very light sterile neutrino corresponding to the mass-squared difference in ranges $10^{-4} - 0.1$ eV$^2$ is expected to be consistent with cosmological mass bounds. It was suggested in ref. \cite{deHolanda:2010am} that the existence of a very light ($\approx 10^{-5} eV^{2}$) sterile neutrino can provide the explanation for the lack of upturn in the solar neutrino oscillation probability below $\approx$ 8 MeV. A recent study has probed the possibility of alleviating the tension between the results of the ongoing beam experiments, T2K and NO$\nu$A for the value $\delta_{cp}$ using very light sterile neutrino with a wide mass difference range of $10^{-5}:0.1$ eV$^2$\cite{deGouvea:2022kma}.

    We focus our study on only one sterile neutrino added to the three light neutrinos, namely the 3+1 framework, and consider a wide mass range for $|\Delta_{41}|$ varying in the range of $10^{-4} - 1$ eV$^2$. The cosmological constraints on the sum of all the neutrino masses imply that the sign of $\Delta_{41}$ can not be negative for $\Delta_{41} >0.1$ eV$^{2}$. However, for lower mass squared differences both signs of $\Delta_{41}$ are possible. In this work, we investigate the possibility of  determining (i) the sign of $\Delta_{31}$ in the presence of a sterile neutrino corresponding to a) $\Delta_{41}=1$ eV$^2$, b)$\Delta_{41}$ in the range of $10^{-4}-0.1$ eV$^2$; (ii) the sign of $\Delta m^{2}_{41}$ for the mass range $10^{-4}- 0.1$ eV$^2$. To answer these questions we use a liquid argon time projection chamber (LArTPC) capable of detecting both beam and atmospheric neutrinos. The typical baseline we have used for the beam neutrinos is $\sim$ 1300 km which is similar to the DUNE experiment. We delineate the sensitivities to  mass ordering by performing a combined analysis of beam and atmospheric neutrinos, along with a separate study for each. Additionally, we present the results including the charge tagging capability of muon capture in liquid argon allowing one to differentiate between $\mu^+$ and $\mu^-$ events in the context of atmospheric neutrinos.      

    The mass ordering in the presence of a light sterile neutrino has been studied in ref. \cite{Thakore:2018lgn} with the additional mass squared difference varying in a wide range in the context of a magnetized iron calorimeter detector proposed by the India-based Neutrino Observatory (INO) collaboration. Recently the sensitivity of the sterile mass ordering in the same mass range has been studied in reference \cite{Chattopadhyay:2022hkw} in the context of the DUNE experiment  using beam neutrinos. We perform our study in the context of a liquid argon time projection chamber detector as in DUNE using both beam and atmospheric neutrino events separately as well as in a combined analysis.    
    
    The plan of the paper is as follows. In section \ref{subsec:3+1}, we present the 3+1 framework which is used for the analysis. In section \ref{subsec:Prob}, we discuss the probability level study of $P_{\mu e}$, $P_{\mu\mu}$ in the presence of a sterile neutrino and explore the effect of sterile mixing and point out where these effects will be significant. Simulation procedures used for the neutrinos coming from the beam and atmosphere, detector specification, and numerical analysis are given in section \ref{subsec:sim}. Next, in section \ref{subsec:res}, we present and discuss the results. Finally, we conclude in section \ref{subsec:con}.

\section{The 3+1 framework:}
\label{subsec:3+1}
The minimal scheme postulated to explain the LSND and  MiniBooNE results is the addition of one light sterile neutrino of mass of the order of eV scale to the three active neutrinos in the SM\cite{Goswami:1995yq,ref:sterile,Donini:2001xp,ref:sterile,LSND:2001aii,Donini:2001xp,Meloni:2010zr,Abazajian:2012ys,Kopp:2013vaa}. In this case, there is one additional independent mass squared difference which we take as $\Delta_{41} = {m^{2}_{4}} - {m^{2}_{1}}$. 
    
\begin{figure}[H]
        \centering
        \includegraphics[width=0.96\linewidth]{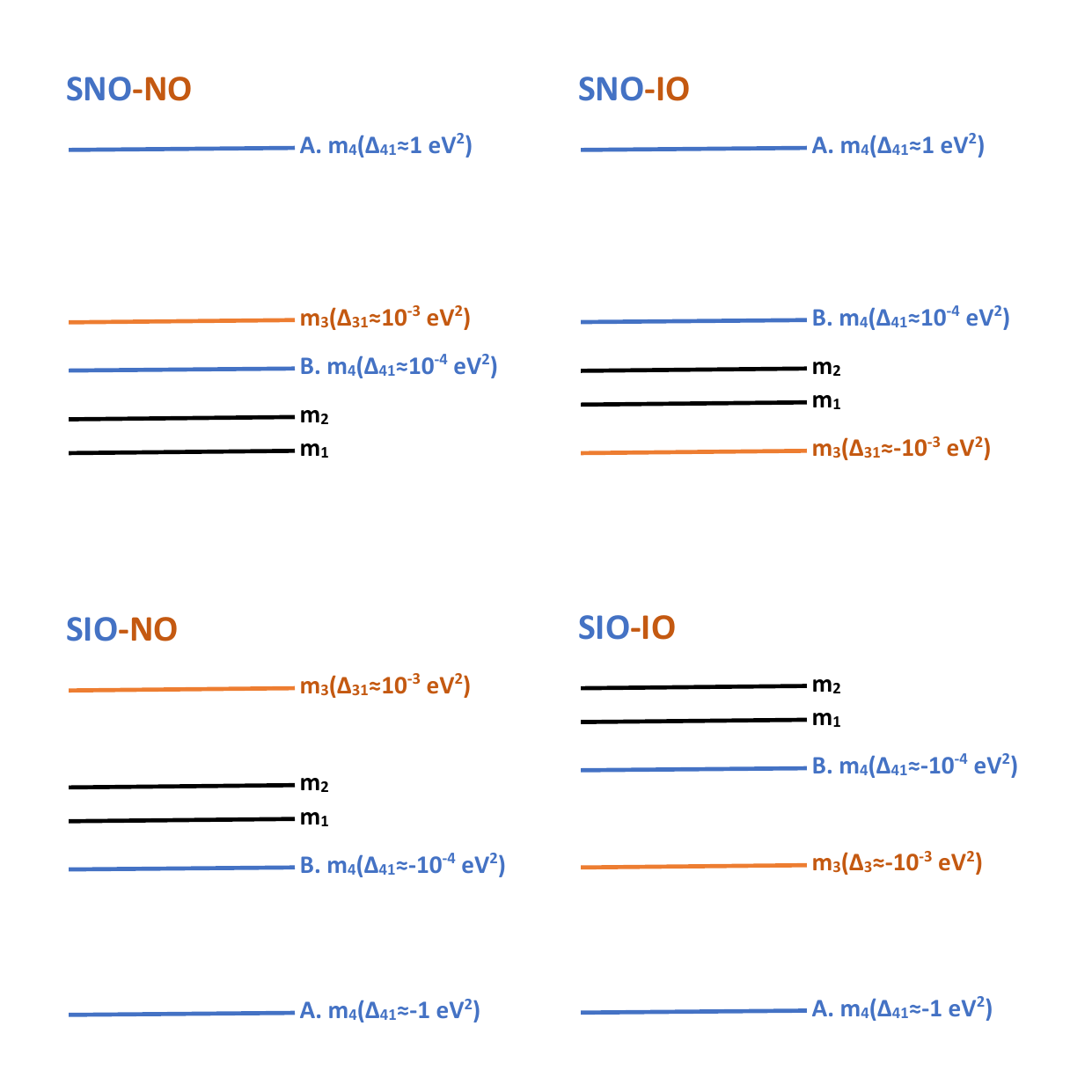}
        \caption{The 3+1 mass spectrum: ordering of mass states in the presence of an extra sterile neutrino state $m_4$ (blue) corresponding to two different sterile mass squared difference: A. $|\Delta_{41}|\sim1$ eV$^2$, B. $|\Delta_{41}|\sim10^{-4}$ eV$^2$ when the standard mass ordering $\Delta_{31}$ lead by $m_3$ (red) can be both +ve and -ve.}
        \label{fig:ster_mass}
    \end{figure}

The possible mass orderings, in this case, are shown in fig. \ref{fig:ster_mass}. We will refer to the sign of $\Delta_{31}$ as the atmospheric mass ordering (AMO) and the sign of $\Delta_{41}$ as sterile mass ordering (SMO) throughout the paper. NO(IO) and SNO(SIO) will signify normal(inverted) ordering for AMO and SMO respectively. There can be four possibilities as follows,\\ 
      (i) SNO-NO: where $\Delta_{41} > 0 $ and $\Delta_{31} > 0 $. The positioning of the 4th state depends on the value of $\Delta_{41}$. For $\Delta_{41} > 10^{-3}$ eV$^2$ the 4th state lies above the 3rd state while if $\Delta_{41}< 10^{-3}$ eV$^2$ it lies below the 3rd state.\\ 
      (ii) SNO-IO: in this case $\Delta_{41} > 0 $ and $\Delta_{31} < 0$ corresponding to inverted ordering of the light active neutrino. The 4th state lies above the three active states with positioning depending on the value of $\Delta_{41}$.\\
      (iii) SIO-NO: this signifies to $\Delta_{41} < 0 $ and $\Delta_{31} > 0$. The 4th state will always lie below the lightest active states with the placement depending on the value of $\Delta_{41}$.\\
      (iv) SIO-IO: for this case both $\Delta_{41} $ and $\Delta_{31}$ are $< 0$. For $\Delta_{41}<10^{-3}$ eV$^2$, the 4th state lies above $m_3$.\\

    Note that the usual 3+1 picture corresponds to the cases (i) and (ii)  with $\Delta m^2_{41} \sim$ eV$^2$. The Cases (iii) and (iv) with $\Delta m^2_{41} \sim$ eV$^2$ are disfavored from cosmology.
     
    For the $ 3+1 $ oscillation framework, the new mixing matrix $ U $ will have three additional mixing angles $ \theta_{14},\theta_{24},\theta_{34} $ corresponding to mixing between the extra light sterile neutrino $ \nu_s $ and the active sector neutrinos along with two new CP phases $ \delta_{14},\delta_{34} $ and can be expressed as follows,
    \begin{align}
         U&=\tilde{R}_{34}(\theta_{34},\delta_{34})R_{24}(\theta_{24})\tilde{R}_{14}(\theta_{14},\delta_{14})\times\nonumber\\&R_{23}(\theta_{23})\tilde{R}_{13}(\theta_{13},\delta_{13})R_{12}(\theta_{12})
    \end{align}\label{eq:U-st}
    where $ \tilde{R}_{ij}=U^\delta_{ij}(\delta_{ij})R_{ij}(\theta_{ij})U^{\dagger\delta}_{ij}(\delta_{ij}) $, $R_{ij}(\theta_{ij})$'s are rotational matrices in i-j plane and $U^\delta_{ij}= diag(1,1,1,e^{\iota\delta_{ij}}) $ with $\delta_{ij}$'s being the CP phases.

    The global analysis of data performed in ref.~\cite{Gariazzo:2017fdh} gives the best-fit values and sterile mixing angles for $\Delta_{41}=1.3$ eV$^2$ as given in Table \ref{tab:3sigma-mixing},
    \begin{table}[H]
        \centering
        \begin{tabular}{|c|c|c|}
         \hline Parameters & $3\sigma$ range & Best Fit\\  \hline $\sin^2 2\theta_{14}$ & 0.04 - 0.09 & 0.079\\  $\theta_{14}$ & $5.76^\circ-8.73^\circ$ & $8.15^\circ$\\\hline $\sin^2 \theta_{24}$ & $6.7 \times 10^{-3}-  0.022 $ & 0.015\\ $\theta_{24}$ & $4.68^\circ - 8.6^\circ$ & $7.08^\circ$\\ \hline
        \end{tabular}
        \caption{$3\sigma$ Levels and Best fit values extracted from \cite{Gariazzo:2017fdh}}
        \label{tab:3sigma-mixing}
    \end{table}
Ref.~\cite{Gariazzo:2017fdh} provides an upper bound of $\theta_{34}\leq 7.4^\circ$ at the $90\%$ C.L. However, the analysis performed in \cite{MINOS:2020iqj} including the MINOS+ data disfavoured the allowed regions in $\theta_{24}$ from above with a new bound at 90\% C.L. $\sin^2 \theta_{24} \leq 0.006$, i.e., $\theta_{24} \leq 4.5^\circ$. Also, the analysis\cite{MINOS:2020iqj} of DayaBay and Bugey3 gives at 90\% C.L. $\sin^2 2\theta_{14} \leq 0.046$. i.e., $\theta_{14} \leq 6.2^\circ$. The bounds on these sterile mixing angles vary with $\Delta_{41}$. For values of $\Delta_{41}$ in the range $10^{-4}:0.1$ eV$^2$, the 90\% C.L. bounds vary from  $\theta_{14}\leq 25^\circ : \theta_{14} \leq 2^\circ$, and $\theta_{24}\leq 56^\circ : \theta_{24} \leq 4^\circ$.
	
The effective interaction Hamiltonian in matter for the 3+1 framework in flavour basis is given as follows,
	\begin{align}\label{eq:H-innt}
		H_{int}&=diag(V_{CC},0,0,-V_{NC})\nonumber\\&=diag(\sqrt{2}G_F N_e,0,0,\sqrt{2}G_F N_n/2),
	\end{align}
where $V_{CC} =\sqrt{2}G_F N_e $ is the charge current interaction potential, $V_{NC} =-\sqrt{2}G_F N_n/2 $ is the neutral current interaction potential and $G_F$ is the Fermi coupling constant with $ N_e, N_n $ corresponding to the density of electron and neutron respectively of the medium in which neutrinos travel.
In the presence of matter, the total Hamiltonian in flavour basis is expressed as follows,
	\begin{equation}\label{eq:H-tot-st}
	    H=\frac{1}{2E_\nu}U\begin{bmatrix}
		            0&0&0&0\\0&\Delta_{21}&0&0\\0&0&\Delta_{31}&0\\0&0&0&\Delta_{41}
		            \end{bmatrix}U^\dagger + \frac{1}{2E_\nu}\begin{bmatrix}
		            A&0&0&0\\0&0&0&0\\0&0&0&0\\0&0&0&\frac{A}{2}
		            \end{bmatrix}
	\end{equation}
where the propagation medium has been considered to be the Earth's matter with neutron density being equal to electron density, i.e, $N_e=N_n$ and the matter potential term is $A=2\sqrt{2}G_F N_e E_\nu$ for neutrino with energy $E_\nu$.

\section{Oscillation Probabilities in presence of a sterile neutrino}
\label{subsec:Prob}
In this section, we present both the appearance probability $P_{e\mu}$ and disappearance probability $P_{\mu \mu}$ at baselines of 1300 km and 7000 km for different sterile parameters. As in the three flavour framework, there are two useful approximate methods to calculate the oscillation probability in the matter: (i) $\alpha-s_{13}$ approximation\cite{Chattopadhyay:2022hkw}, (ii) $\Delta_{21}=0$ approximation\cite{Chatterjee:2022pqg}. In the first case $\alpha=\frac{\Delta_{21}}{\Delta_{31}}, \sin{\theta_{13}}$ are considered small parameters whereas in the second case $\Delta_{21}$ is considered as negligible.
The probabilities evaluated in these approximations considering $\theta_{34}=0^\circ$ are presented in the following section. 

\begin{widetext}
\subsection{$\alpha-s_{13}$ Approximation}
The appearance probability calculated using $\alpha-s_{13}$ approximation in \cite{Chattopadhyay:2022hkw} is well suited for any $\Delta_{41}$ values and baseline of 1300 km and can be expressed as,
\begin{equation}\label{eq:Pme_alpha}
    \begin{split}
       P_{\mu e}^{m}&=4 s_{13}^2 s_{23}^2 \frac{\sin^2 [(A'-1)\Delta]}{(A'-1)^2}+ 8 \alpha s_{13}s_{12}c_{12}s_{23}c_{23} \frac{\sin[A^\prime \Delta]}{A'} \frac{\sin [(A'-1)\Delta]}{A'-1} \cos(\Delta+\delta_{13})\\
    +&4 s_{13} s_{14} s_{24} s_{23}\frac{\sin [(A'-1)\Delta]}{A^\prime-1} [P_{14}^{s}\sin{\tilde{\delta}_{14}}+ P_{14}^{c}\cos{\tilde{\delta}_{14}}] 
    \end{split},
\end{equation}
where the terms corresponding to sterile neutrino are,
\begin{equation}\label{eq:pme_14_s}
    \begin{split}
        P_{14}^{s} = R[\frac{1}{2}A'c_{23} + (R-1)(1+s_{23}^2)]\frac{\sin[(R-1+\frac{A'}{2})\Delta]}{R-1+\frac{A'}{2}} \frac{\sin[(R-\frac{A'}{2})\Delta]}{R-\frac{A'}{2}} + R c_{23}^2 \sin[(R-1-\frac{A'}{2})\Delta]\frac{\sin[(R+\frac{A'}{2})\Delta]}{R+\frac{A'}{2}}
    \end{split}
\end{equation}
\begin{equation}\label{eq:pme_14_c}
    \begin{split}
        P_{14}^{c}&=\frac{R}{R-\frac{1}{2}}\left([R-\frac{1}{2 }s_{23}^{2} -\frac{1}{2}] \cos[(R-1-\frac{A'}{2})\Delta] \frac{\sin[(R-\frac{A'}{2})\Delta]}{R-\frac{A'}{2}} + s_{23}^{2} (R-1)\cos[(R-\frac{A'}{2})\Delta]\frac{\sin[(R-1+\frac{A'}{2})\Delta]}{R-1+\frac{A'}{2}}\right. \\&\left.+ s_{23}^{2} \frac{\sin[(A'-1)\Delta]}{A'-1}\right) + R c_{23}^{2} \cos[(R-1+\frac{A'}{2})\Delta] \frac{\sin[(R+\frac{A'}{2})\Delta]}{R+\frac{A'}{2}}
    \end{split},
\end{equation}
where $A'=\frac{A}{\Delta_{31}}$, $R=\frac{\Delta_{41}}{\Delta_{31}}$, $\Delta=\frac{1.27 \Delta_{31}L}{E}$, $\tilde{\delta}_{14}=\delta_{13}+\delta_{14}$ and $c_{ij}\sim \cos \theta_{ij}, s_{ij} \sim \sin \theta_{ij}.$\\
at limit $R>>1$, for $R>>\frac{A'}{2}$, approximately $R-\frac{A'}{2}\simeq R+\frac{A'}{2}\simeq R$, also $R-\frac{1}{2}\simeq R$, $R-\frac{1}{2 }s_{23}^{2} -\frac{1}{2}\simeq R$

\begin{eqnarray}\label{eq:p14s-p14c-r>1}
    P_{14}^s \simeq & \frac{1}{2}A'c_{23}\frac{\sin[(R-1)\Delta]}{R-1} \sin[R\Delta] +  2\sin[(R-1)\Delta]\sin[R\Delta] \simeq (\frac{1}{2R}A'c_{23}+2) \sin[R\Delta]^2\\
    P_{14}^{c} \simeq & (1+c_{23}^{2}) \cos[(R-1)\Delta] \sin[R\Delta] + s_{23}^{2}\cos[R\Delta]\sin[(R-1)\Delta] + s_{23}^{2} \frac{\sin[(A'-1)\Delta]}{A'-1} \simeq \sin [2R\Delta] + s_{23}^{2} \frac{\sin[(A'-1)\Delta]}{A'-1}
\end{eqnarray}
\subsection{$\Delta_{21}=0$ Approximation}
The dominant term in the $\nu_\mu-\nu_e$ oscillation probability in OMSD approximation, valid for $\Delta_{21} L/E << 1$, e.g., at 7000 km baseline around the resonance energy (7 GeV), is given by \cite{Chatterjee:2022pqg},
    \begin{equation}\label{eq:Pme1-omsd}
        \begin{split}
         P_{\mu e}^1&=4\cos^2\theta_{13m}\cos^2\theta_{14m}\sin^2\theta_{13m}(\cos^2\theta_{24m}\sin^2\theta_{23}-\sin^2\theta_{14m}\sin^2\theta_{24m})\sin^2\frac{1.27\Delta_{31}^m L}{E}\\
         &+2\cos^3\theta_{13m}\cos^2\theta_{14m}\sin\theta_{13m}\sin\theta_{14m}\sin 2\theta_{24m}\sin\theta_{23}\sin{\frac{1.27\Delta_{31}^m L}{E}}\sin(\frac{1.27\Delta_{31}^m L}{E}+\delta_{13}-\delta_{14})\\
         &-2\cos\theta_{13m}\cos^2\theta_{14m}\sin^3\theta_{13m}\sin\theta_{14m}\sin 2\theta_{24m}\sin\theta_{23}\sin{\frac{1.27\Delta_{31}^m L}{E}}\sin(\frac{1.27\Delta_{31}^m L}{E}-\delta_{13}+\delta_{14})
        \end{split}
    \end{equation}
where $\theta_{13m},\theta_{14m},\theta_{24m}$ are modified mixing angles, $\Delta_{ij}^m$'s are modified mass-squared differences as defined in ref. \cite{Chatterjee:2022pqg}.
\end{widetext}

\subsection{Effect of sterile parameters on sign of $\Delta_{31}$}
In this section, we discuss the dependence of the probabilities on the sterile parameters. For our study, we chose two illustrative mixing angles $\theta_{14} = \theta_{24}= 4^\circ$ and/or $7^\circ$. The phases $\delta_{13}$ and $\delta_{14}$ are varied in the range $-180^\circ$ to $180^\circ$, unless otherwise mentioned. The probabilities for both $\theta_{34}=0^\circ$ and non-zero $\theta_{34}=7^\circ$ are discussed. In this section, GLoBES\cite{Huber:2004ka} package is used to generate the probabilities.
\subsubsection{Effect of non-zero $\theta_{14},\theta_{24}$}
    \begin{figure}[h!]
		\centering
		\includegraphics[width=0.96\linewidth]{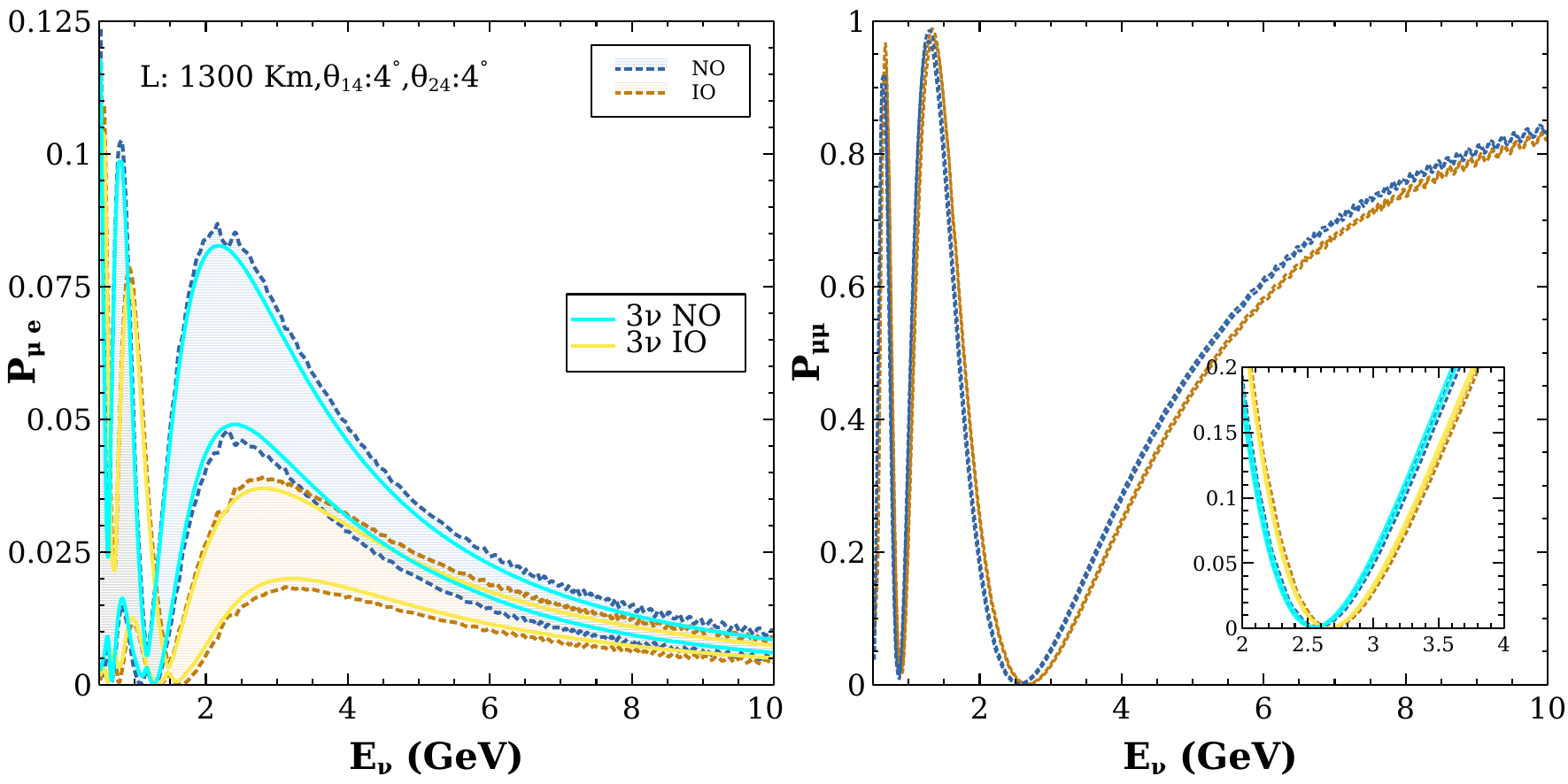}
            \includegraphics[width=0.96\linewidth]{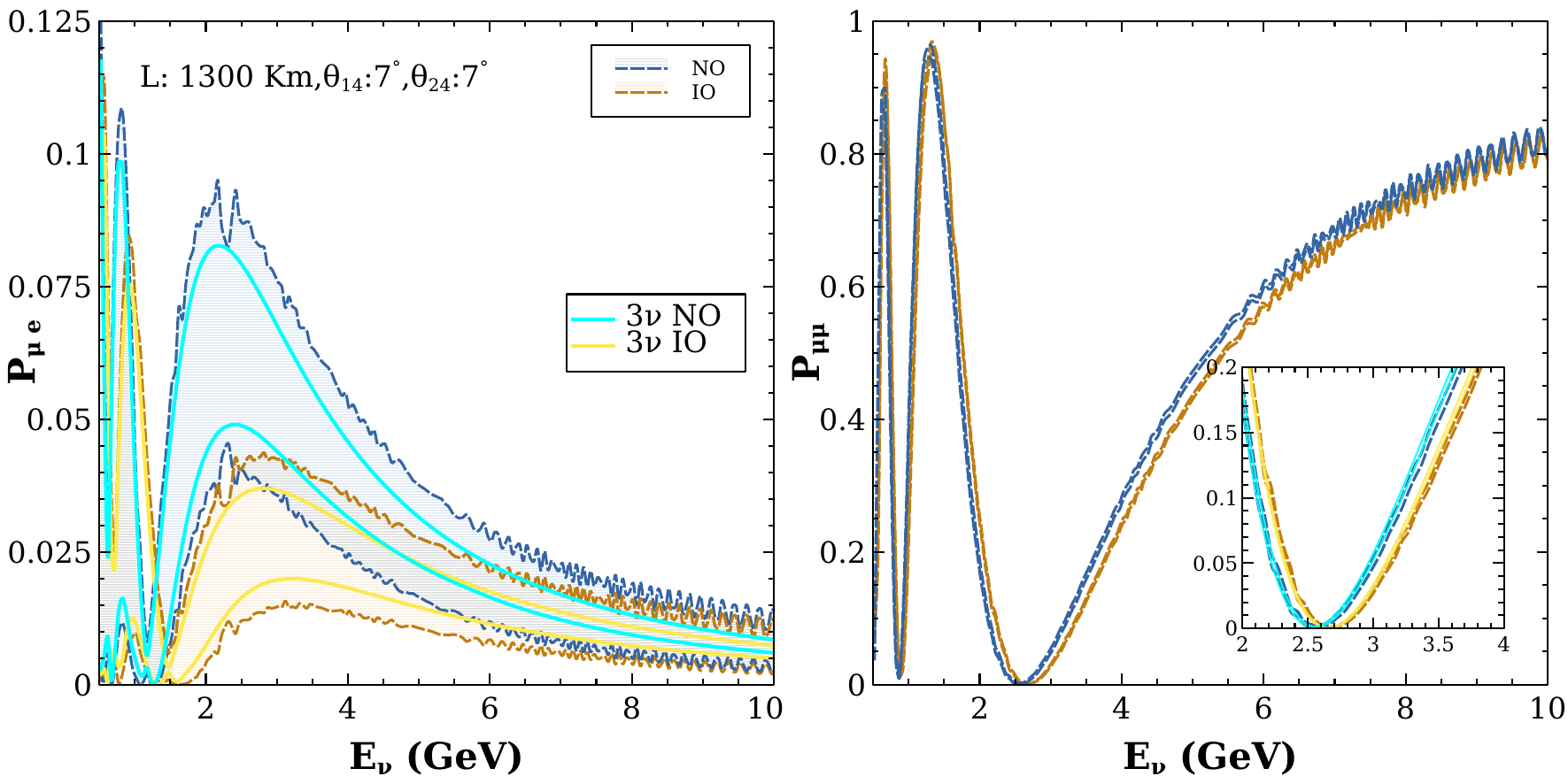}
		\caption{Probabilities $P_{\mu e}$(left) and $P_{\mu\mu}$ (right) as a function of energy $E_\nu$ due to variation of phases $\delta_{13},\delta_{14}$ for NO and IO at 1300 km baseline for $\Delta_{41}=1$ eV$^2$. Blue and orange bands in top (bottom) panels refer to varied phases for $\theta_{14}=4^\circ$ ($7^\circ$), $\theta_{24}=4^\circ$ ($7^\circ$) corresponding to NO and IO respectively. The regions between cyan(yellow) curves are due to the variation of $\delta_{13}$ in $3\nu$ case for NO(IO).}
		\label{fig:pme-pmm_hr_1300}
    \end{figure}
We have plotted the appearance (left), and disappearance (right) probabilities in fig.~\ref{fig:pme-pmm_hr_1300} as a function of the neutrino energy varying the phases $\delta_{13},\delta_{14}$. The blue[NO] and orange[IO] bands at the top (bottom) panels refer to mixing angles $\theta_{14}=\theta_{24}=4^\circ(7^\circ)$ in the 3+1 framework. Whereas, the regions between cyan and yellow lines correspond to the variation of $\delta_{13}$ in three generation framework in NO and IO respectively. In the right panel, we also show $P_{\mu\mu}$ over $2-4$ GeV in a magnified inset. The important observations are as follows,
    \begin{itemize}

        \item In the 3+1 framework, the probability bands corresponding to NO and IO are closer than those in the three generation framework. This suggests a reduced hierarchy sensitivity in the 3+1 framework.

        \item The gap between NO and IO bands increases with the decreasing values of the sterile mixing angles.
        
        \item In $P_{\mu e}$ channel, for the 3+1 framework the difference between the two probability bands is seen in the energy range 1-3 GeV.

        \item The disappearance channel probability $P_{\mu \mu}$ doesn't depend on the phases, as can be seen from the narrow bands in NO and IO cases, for both three and 3+1 frameworks.
         
         \item The $P_{\mu \mu}$ curves for opposite mass orderings are hard to separate from each other at energies lower than 2 GeV. However, some demarcation is visible at energies in the range of 2-7 GeV for both three generation and 3+1 generation, as shown in the right panels.
         
         \item In the disappearance channel, we don't see a significant effect of variation of the sterile-active mixing angles $\theta_{14},\theta_{24}$ on the probability bands.
        
    \end{itemize}

    \begin{figure}[H]
        \centering
        \includegraphics[width=0.96\linewidth]{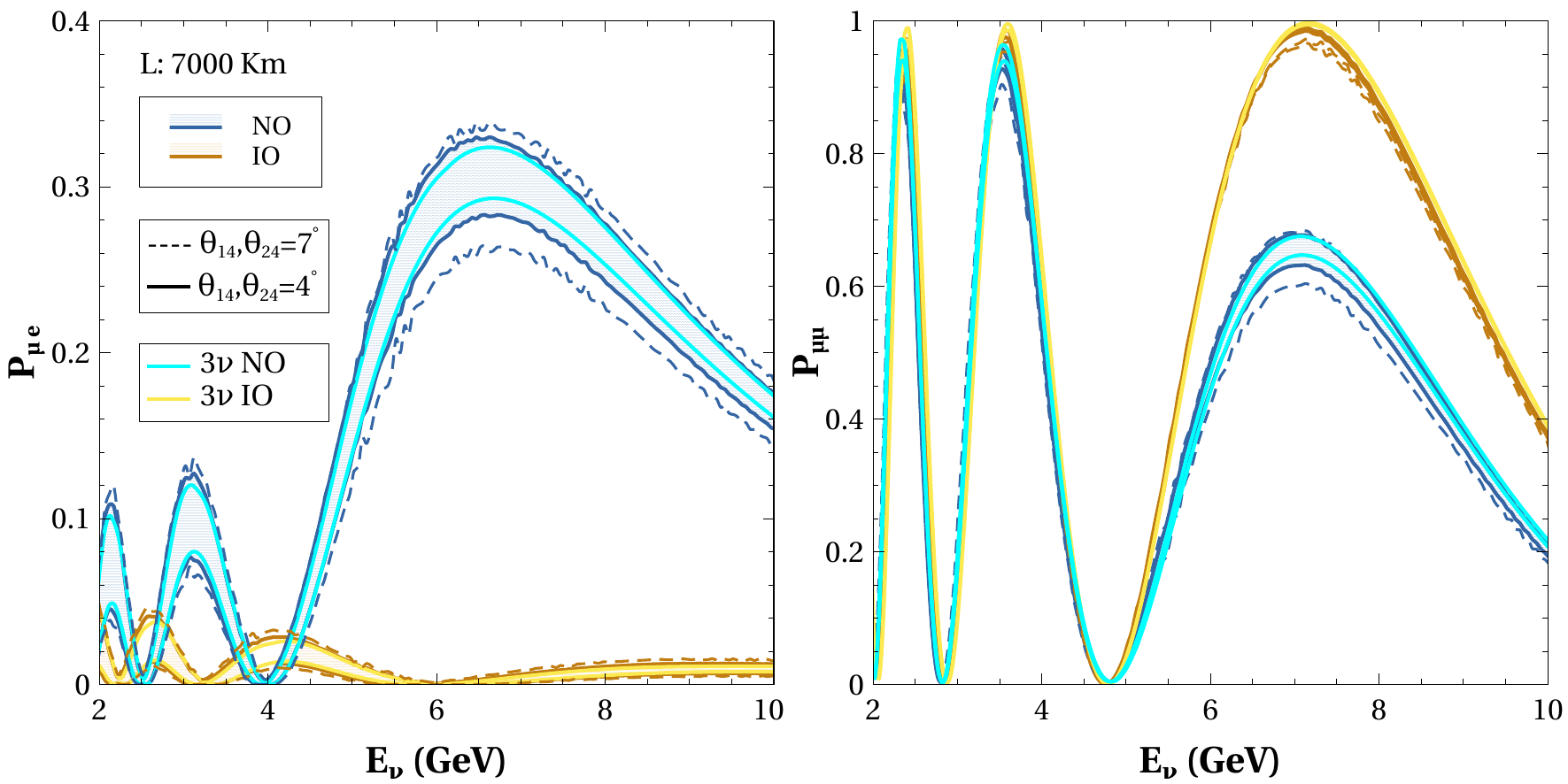}
        \caption{Probabilities $P_{\mu e}$(left) and $P_{\mu\mu}$ (right) as a function of energy $E_\nu$ at 7000 km baseline for $\Delta_{41}=1$ eV$^2$. The shaded bands of blue and orange refer to varied phases $\delta_{13},\delta_{14}$ for $\theta_{14}=4^\circ$ ($7^\circ$) for NO(IO). The regions between cyan(yellow) curves are due to the variation of $\delta_{13}$ in $3\nu$ case for NO(IO).}
        \label{fig:pme-pmm_hr_7k}
    \end{figure}
In fig.~\ref{fig:pme-pmm_hr_7k}, the appearance (left) and disappearance (right) probabilities have been shown at 7000 km baseline. Such baselines will be relevant for atmospheric neutrinos. The significant observations are as follows,
    \begin{itemize}
        \item There is a prominent difference between the probability bands of NO and IO, implying sensitivity to mass ordering  at 7000 km baseline in the 3+1 framework. The difference decreases for sterile case w.r.t. the standard one.
        
        \item Also, with lower values of $\theta_{14},\theta_{24}$, the gap between the probability bands of opposite mass ordering increases.
        
        \item For $P_{\mu\mu}$ channel, a significant gap exists between bands of opposite mass orderings at energies higher than 4 GeV whereas, in $P_{\mu e}$ channel, the gap is present even at much lower energies of 3 GeV.

        %\item From $P_{\mu e}$ plot, at 7 GeV the gap between NO and IO bands is around $0.23$ which is similar to what has been calculated using \eqref{eq:del_P_hr_np}.
    \end{itemize}
In order to understand the features of the fig.~\ref{fig:pme-pmm_hr_1300}, ~\ref{fig:pme-pmm_hr_7k} from analytical expression, we compute the minimum difference in the probability $P_{\mu e}$ for two different mass orderings by varying the phases,
    \begin{equation}
	    \Delta P \equiv [P_{\mu e}^{1NO}(\delta_{13}^{NO},\delta_{14}^{NO})- P_{\mu e}^{1IO}(\delta_{13}^{IO},\delta_{14}^{IO})]_{min}
    \end{equation}
where the other parameters have been kept fixed. Using only the dominant first term of \eqref{eq:Pme1-omsd}, the difference in probability can be expressed as,
    \begin{equation}\label{eq:del_P_hr}
	    \begin{split}
	        \Delta P &=\Delta P_{np}+\\& A_1[\sin^2(M-N)\cos\delta^{NO} - \sin^2(M+N)\cos\delta^{IO}]+\\& A_2 [\sin 2(M-N)\sin\delta^{NO} + \sin 2(M+N)\sin\delta^{IO}]
	    \end{split}
    \end{equation}	
where $\Delta P_{np}$ is the part with no phases involved, and  is given as follows,
    \begin{align}\label{eq:del_P_hr_np}
	   \Delta P_{np} &=\cos^2\theta_{14m}\sin^2 2\theta_{13m}(\sin^2\theta_{24m}\sin^2\theta_{14m}-\nonumber\\ &\cos^2\theta_{24m}\sin^2\theta_{23}) \sin 2M \sin 2N
    \end{align}
The other part containing the phases depends on the amplitude parameters $A_{1}, A_{2}$ and the frequency parameters $M, N$ that are defined as, 
    \begin{gather}
        A_1=\cos^2\theta_{14m}\cos2\theta_{13m}\sin 2\theta_{13m}\sin\theta_{14m}\sin2\theta_{24m}\sin\theta_{23}\\
	A_2=\cos^2\theta_{14m}\sin 2\theta_{13m}\sin\theta_{14m}\sin2\theta_{24m}\sin\theta_{23}
    \end{gather}
    \begin{align}
	M&=\Delta_{31}\cos 2(\theta_{13}-\theta_{13m})\times \frac{1.27L}{E}\\
	N&=A\cos2\theta_{13m}(1+\cos^2\theta_{14}+\cos^2\theta_{14}\sin^2\theta_{24})\times \frac{1.27L}{E}
    \end{align}

We use $\Delta_{31}=2.5\times 10^{-3}$ eV$^2$, $\theta_{24}=7^{\circ}$, $\theta_{14}=7^{\circ}$, $\theta_{23}=45^\circ$ to calculate the $\Delta P$ at the first oscillation maxima. For $L=1300$ km at $E= 2.5$ GeV, $\Delta P < 0$, which means the minima of the NO curve is below the maxima of the IO curve. This implies the overlap of NO and IO bands suggesting the presence of degeneracy. Whereas, for 7000 Km at $E=7$ GeV, $\Delta P>0$, i.e., mass ordering can be determined even with varying the phases $\delta_{13}$ and $\delta_{14}$.

\subsubsection{Effect of non-zero $\theta_{34}$}
    \begin{figure}[H]
		\centering
		\includegraphics[width=0.96\linewidth]{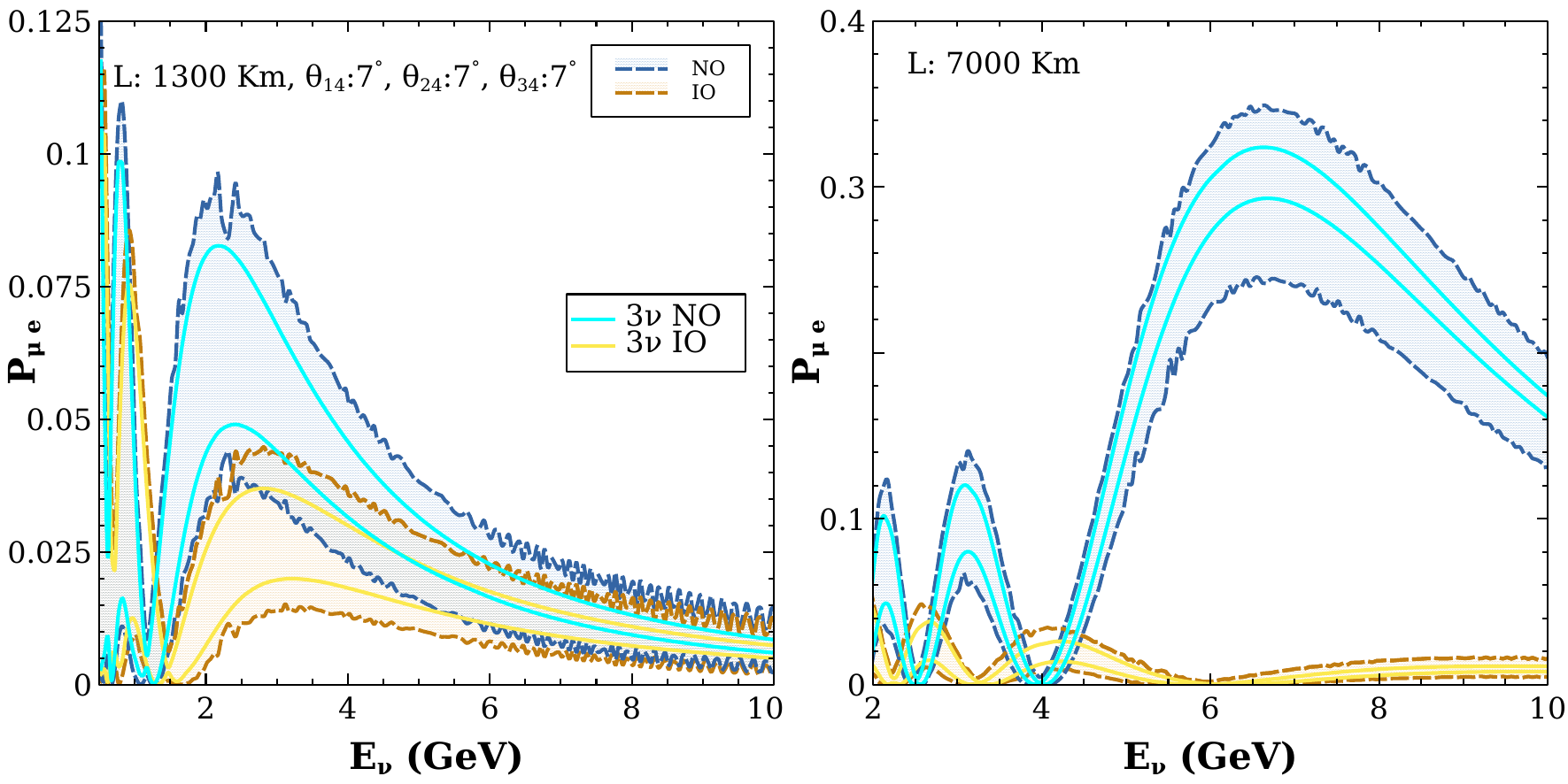}
		\caption{$P_{\mu e}$ as a function of energy $E_\nu$ with the variation of phases $\delta_{13},\delta_{14},\delta_{34}$ for NO (blue) and IO(orange), $\Delta_{41}=1$ eV$^2$ at 1300 km (left) and 7000 km (right). Shaded blue (orange) bands refer to $\theta_{14},\theta_{24},\theta_{34}=7^\circ$ for NO(IO). Regions between cyan(yellow) curves are due to variation of $\delta_{13}$ in $3\nu$ case for NO(IO).}
		\label{fig:pme-hr_7k,dune-th34}
    \end{figure}
In fig.~\ref{fig:pme-hr_7k,dune-th34}, the appearance probability is plotted as a function of neutrino energy at 1300 km (left) and 7000 km (right) baseline for $\theta_{14},\theta_{24}, \theta_{34}=7^\circ$. The shaded blue (yellow) bands are due to the variation of the phases $\delta_{13},\delta_{14},\delta_{34}$ for NO(IO). The regions between the solid cyan(yellow) curves correspond to the variation of $\delta_{13}$ in the 3$\nu$ framework for NO(IO). The most important observation is a notable decrease in the gap between NO and IO bands at both baselines. In the case of 7000 km, there is still a gap between the opposite mass ordering bands, which gets diminished for a non-zero $\theta_{34}$.
    
\subsubsection{Effect of $|\Delta_{41}|$}
In this section, the sensitivity to atmospheric mass ordering is studied at the probability level, with the sterile mass squared difference $\Delta_{41}$ in the range $10^{-4}$  - $0.1$ eV$^2$.

To understand the effect of $\Delta_{41}$ on sensitivity to atmospheric mass ordering, we probe the difference in the appearance probability $\Delta P_{\mu e}$ as a function of $\Delta_{41}$.
\begin{align}\label{eq:Dpme-mo}
    \Delta P_{\mu e}=|P_{\mu e}^{true}(\Delta_{31},\delta_{13}, \delta_{14}) - P_{\mu e}^{test}(-\Delta_{31}', \delta_{13}', \delta_{14}')|_{min}
\end{align}
We compute the minimum difference $\Delta P_{\mu e}$ (using GLoBES), by considering a particular AMO with constant $\delta_{13},\delta_{14}$ in $P_{\mu e}^{true}$ whereas $P_{\mu e}^{test}$ is calculated for the opposite AMO by varying the phases and $|\Delta_{31}|$. The phases are varied over their full range and for $\Delta_{31}$ variation over the current $3\sigma$ range in the opposite mass ordering is considered.
\begin{figure}[H]
        \centering
        \includegraphics[width=0.96\linewidth]{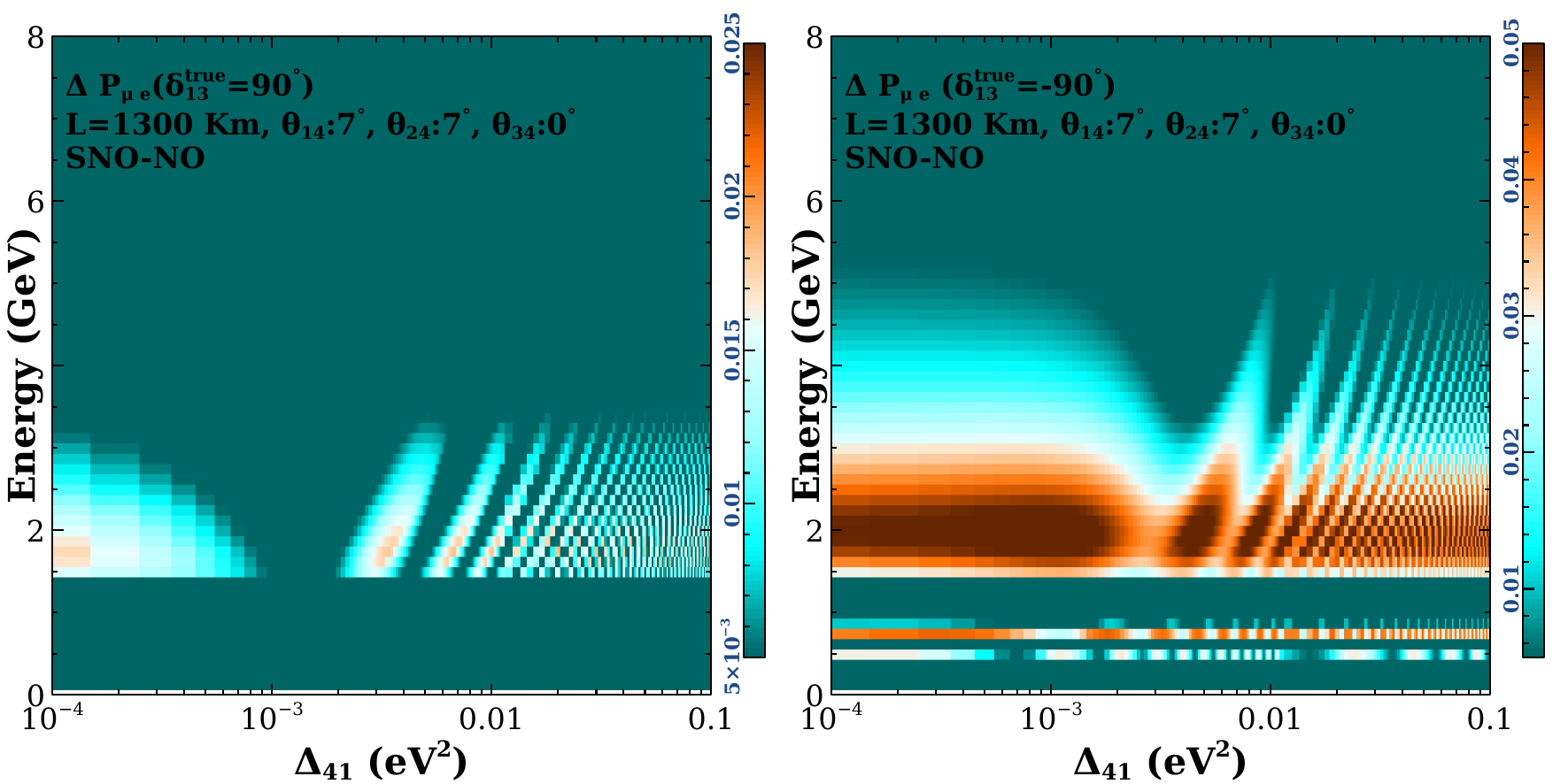}
        \includegraphics[width=0.96\linewidth]{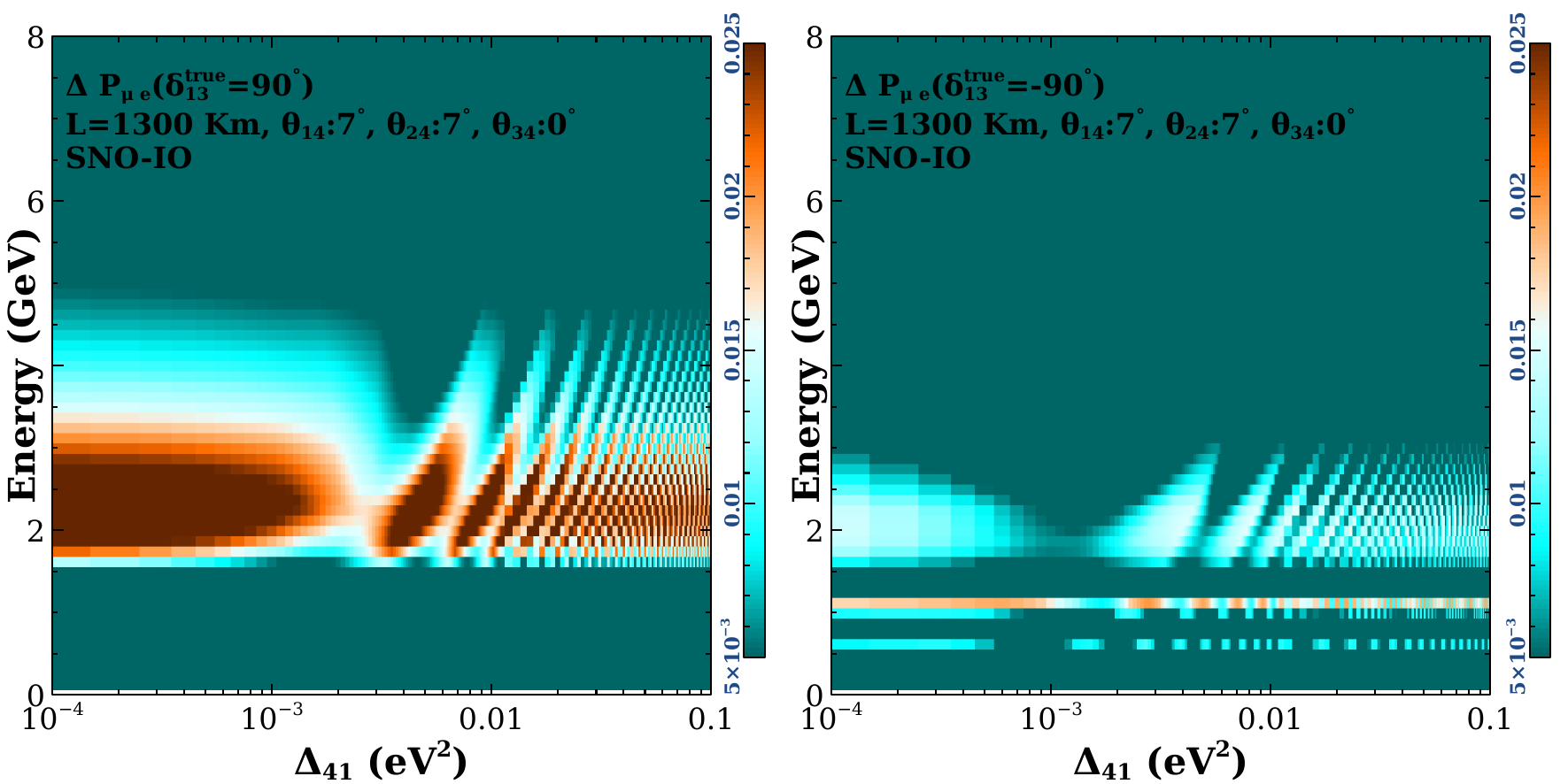}
        \caption{The difference in appearance channel probability $\Delta P_{\mu e}$ for opposite AMO as a function of $\Delta_{41}^{true}$ and $E_\nu$ at 1300 km baseline for SNO-NO (above) and SNO-IO (below). The plots in the left (right) panel has $\delta_{13}^{true}=90^\circ(-90^\circ)$ and fixed $\delta_{14}=0^\circ$.}
        \label{fig:dPme-mo-beam}
    \end{figure}
In fig.~\ref{fig:dPme-mo-beam}, we illustrate the probability difference $\Delta P_{\mu e}$ in the $\Delta_{41}-E_{\nu}$ plane at 1300 km. The important observations are as follows,
\begin{itemize}
    \item Around $\Delta_{41}=2.5\times 10^{-3}$ eV$^2$, $\Delta P_{\mu e}$ is either very high or low depending upon the values of $\delta_{13}$ and NO/IO.

    \item We observe an oscillating pattern of $\Delta P_{\mu e}$ along $\Delta_{41}$ for a fixed energy. This oscillation becomes rapid at higher $\Delta_{41}$ values.

    \item Significant contribution to $\Delta P_{\mu e}$ is seen for energies in the range of $1.5-4$ GeV.

    \item For the SNO-NO case (top panels), the occurrence for maxima and minima reverses for $\delta_{13}=90^\circ$, and $-90^\circ$.

    \item However, for the SNO-IO case, the maxima and minima occur at the same $\Delta_{41}$ for $\delta_{13}=90^\circ,-90^\circ$. Although, the magnitude is higher for $\delta_{13}=90^\circ$.
\end{itemize}
    
\subsection{Effect of the sign of $\Delta_{41}$ in $P_{\mu e}$ channel}
 As we consider $\Delta_{41}$ in the rage $5\times10^{-4}:10^{-1}$ eV$^2$, the sterile mass ordering also becomes unknown giving us four possibilities depending on the ordering of the three active states. The difference in the probability for the opposite signs of $\Delta_{41}$ is defined as,
    \begin{equation}
        \Delta P_{s}= |P_{\mu e,\mu\mu}^{true}(+\Delta_{41},\delta_{13},\delta_{14})- P_{\mu e,\mu\mu}^{test}(-\Delta_{41}',\delta_{13}',\delta_{14}')|_{min}
    \end{equation}\label{eq:dPme-hr-st}
    \begin{figure}[h!]
        \centering
        \includegraphics[width=0.96\linewidth]{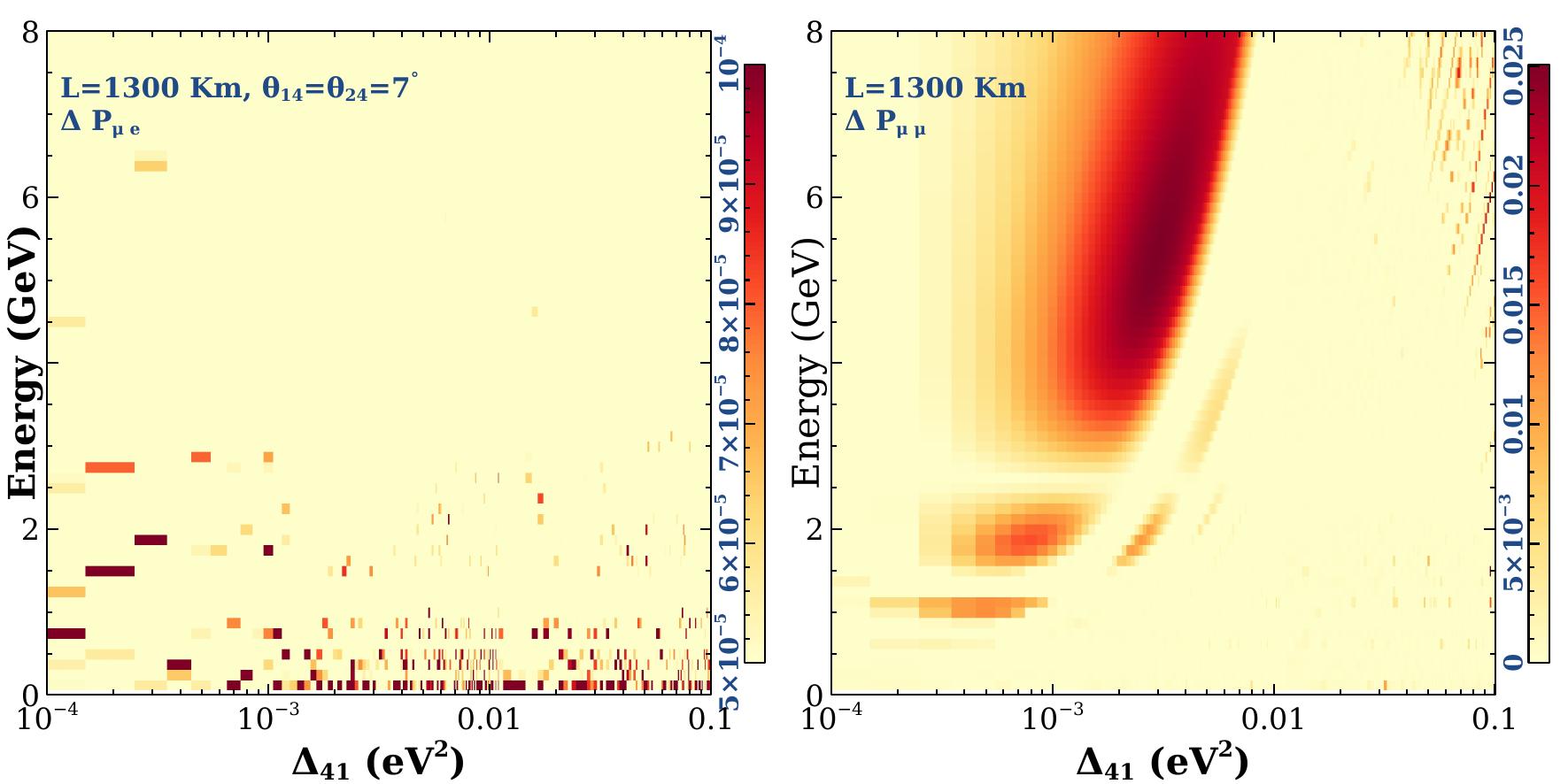}
        \includegraphics[width=0.96\linewidth]{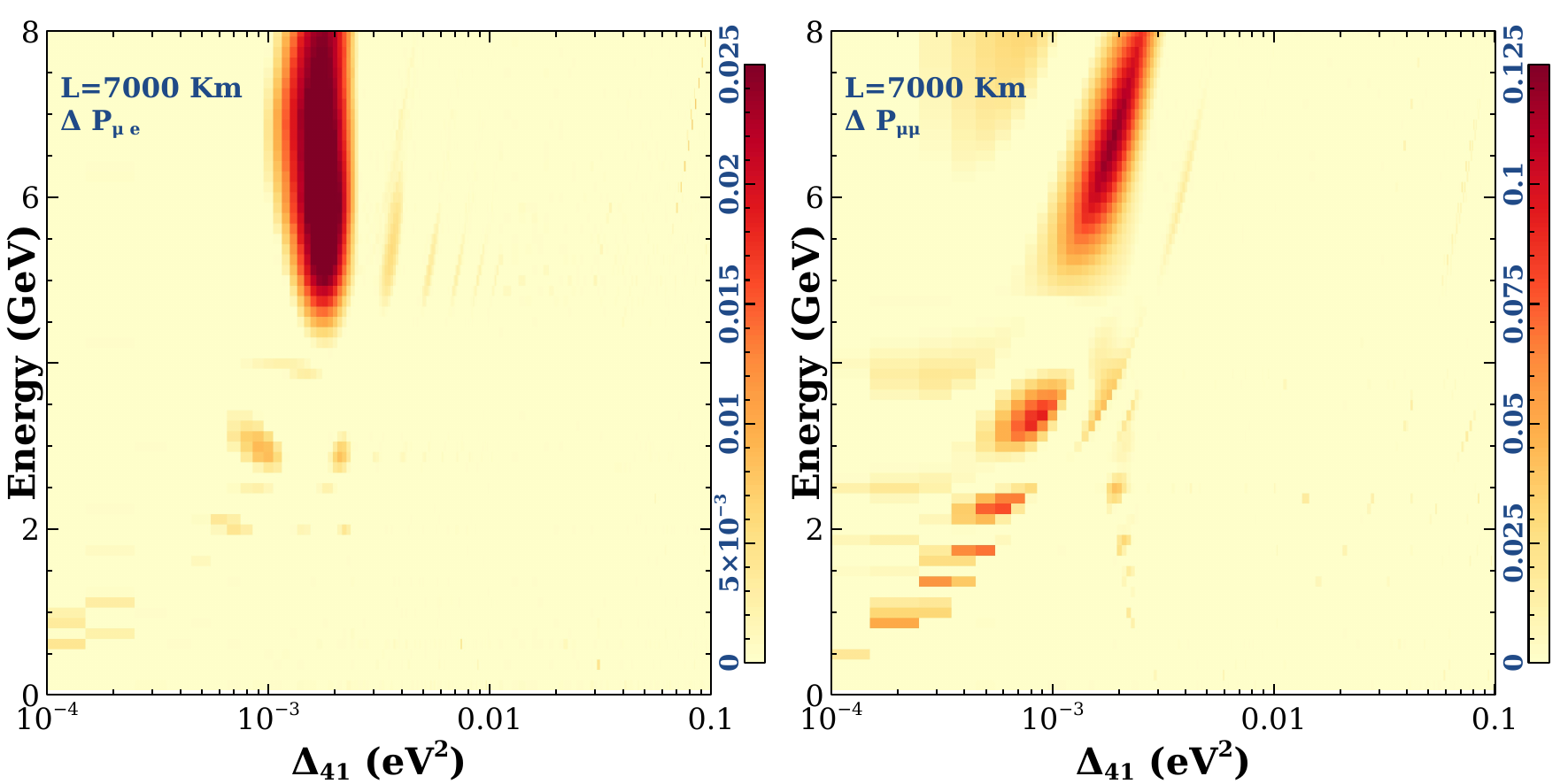}
        \caption{Difference in the appearance probability $\Delta P_{\mu e}$(left), and disappearance probability $\Delta P_{\mu \mu}$(right) for different sterile mass ordering in the $\Delta_{41}-E_\nu$ plane at 1300 km(top), 7000 km(bottom).}
        \label{fig:dPme-st-mh-hr-beam}
    \end{figure}
    
We plot $\Delta P_{s}$ (using GLoBES) by marginalizing over $\Delta_{41}'$ in opposite SMO and phases $\delta_{13}',\delta_{14}'$  in fig.~\ref{fig:dPme-st-mh-hr-beam} for appearance (left) and disappearance channel(right) over a wide range of the sterile mass squared difference and neutrino energy at 1300 km baseline (top) and 7000 km (bottom). We have taken constant values of $\theta_{14},\theta_{24}=7^\circ$, $\delta_{13}=-90^\circ,\delta_{14}=90^\circ$ in $P_{\mu e,\mu\mu}^{true}$. It can be observed that the high values of $\Delta P_{s}$ are mostly concentrated in the mass square range of $10^{-3}:10^{-2}$ eV$^2$ range. In $P_{\mu e}$ channel, the contribution is lower than $P_{\mu \mu}$. The difference is observed to be larger at higher baselines.

In $P_{\mu e}$ channel, probability difference is higher around $\Delta_{41}=1-2\times 10^{-3}$ eV$^2$ while a dip is found immediately after around $\Delta_{41}=2.5-3\times 10^{-3}$ eV$^2$ . In $P_{\mu \mu}$ channel for energy below 4 GeV, a similar pattern is observed. However, at higher energies,  a higher $\Delta P$ value is observed.

\section{Simulation procedure and details of the experimental setup}
\label{subsec:sim}
The experimental setup under consideration consists of a megawatt-scale muon neutrino beam source accompanied by a near detector (ND) and a far detector(FD). The ND will be placed close to the source of the beam, while the FD, comprising a 40 Kton LArTPC detector is placed at a distance of 1300 km away from the neutrino source. The large LArTPC at an underground observatory is also capable of observing atmospheric neutrinos. The proposed DUNE experiment has a similar experimental configuration\cite{DUNE:2016ymp}. In this analysis, both neutrino beam coming from the accelerator and the atmospheric neutrinos have been considered. 

\subsection{Events from accelerator beam}
 A beam-power of 1.2MW leading to a total exposure of $10\times 10^{21}$ pot has been implemented for the numerical analysis. The neutrino beam simulation has been carried out using the GLoBES\cite{Huber:2004ka} software. We assume the experiment to be running for 3.5 years each in the neutrino mode and the antineutrino mode. 

\subsection{Atmospheric events}
The atmospheric neutrinos produce muons and electrons. These event rates corresponding to an energy bin of width $\mathrm{dE}$ and in a solid angle bin of width ${\mathrm{d \Omega}}$ are expressed as,
    \begin{equation} \label{eq:muevent}
    \rm{ \frac{d^2 N_{\mu}}{d \Omega \;dE} = \frac{\sigma_c D_{eff}}{2\pi} \left[\left(\frac{d^2 \Phi_\mu}{d \cos \theta \; dE}\right) P_{\mu\mu} +\left(\frac{d^2 \Phi_e}{d \cos \theta \; dE}\right)P_{e\mu}\right]},
    \end{equation}
   \begin{equation} \label{eq:eevent}
    \mathrm{ \frac{d^2 N_e}{d \Omega \;dE} = \frac{\sigma_c D_{eff}}{2\pi}\left[\left( \frac{d^2 \Phi_{\mu}}{d \cos \theta \; dE}\right)P_{\mu e} + \left(\frac{d^2 \Phi_e}{d \cos \theta \; dE}\right)P_{e e} \right]}
    \end{equation}
Here ${\mathrm{\Phi_{\mu}}}$ and ${\mathrm{\Phi_{e}}}$ stands for the $\nu_\mu$ and ${\mathrm{\nu_e}}$ atmospheric fluxes \cite{PhysRevD.92.023004} respectively, the disappearance and appearance probabilities are given as  $P_{\mu\mu}$ and $P_{e\mu}$ respectively, $\rm{\sigma_c}$ is the total CC cross section and $\rm{D_{eff}}$ is the detector efficiency.
The energy and angular resolutions for the LArTPC detector, implemented using Gaussian resolution function R, are defined as follows,
    \begin{equation} \label{eq:esmear}
    \mathrm{ R_{E}(E_t,E_m) = \frac{1}{\sqrt{2\pi}\sigma} \exp\left[-\frac{(E_m - E_t)^2}{2 \sigma^2}\right]}\,.
    \end{equation}
    \begin{equation} \label{eq:anglesmear}
    \rm{ R_{\theta}(\Omega_t, \Omega_m) = N \exp \left[ - \frac{(\theta_t -\theta_m)^2 + \sin^2 \theta_t ~(\phi_t - \phi_m)^2}{2 (\Delta\theta)^2} \right] } \,,
    \end{equation}
where N is the normalization constant.
Here, $\rm{E_m}$ (${\rm{\Omega_m}}$) and $\rm{E_t}$ (${\rm{\Omega_t}}$), denote the measured and true values of energy (angle) respectively. The smearing width $\sigma$ is a function of $\rm{E_t}$. The smearing function for the zenith angle is a bit more complicated. The direction of the incident (measured) neutrino is specified by two variables: the polar angle ${\rm{\theta_t}}({\rm{\theta_m}})$ and the azimuthal angle ${\rm{\phi_t}}({\rm{\phi_m}})$, denoted together by ${\rm{\Omega_t}}({\rm{\Omega_m}})$. The measured direction denoted by ${\rm{\Omega_m}}$ is expected to be within a cone of half angle $\Delta \theta$ of the true direction.
The far detector (LArTPC) parameters assumed are mentioned in \autoref{table:LAr-parameter}\cite{Barger:2014dfa}.
    \begin{table}[H]
		\centering
		\begin{tabular}{|c|c|}
			\hline
			Parameter Uncertainty & Value \\\hline
			$\mu^{+/-}$ $\Delta \theta$ & $2.5^\circ$\\
			$e^{+/-}$ $\Delta \theta$ & $3.0^\circ$\\
			($\mu^{+/-}$, $e^{+/-}$) Energy & GLB files for each E bin \cite{DUNE:2016ymp} \\ Detection efficiency  & GLB files for each E bin \cite{DUNE:2016ymp}\\
			Flux normalization & 20$\%$\\ Zenith angle dependence & 5$\%$\\
			 Cross section & 10$\%$\\
			Overall systematic & 5$\%$\\
			Tilt & 5$\%$\\
		    \hline
		\end{tabular}
		\caption{Assumptions of the LArTPC far detector parameters and uncertainties.}
		\label{table:LAr-parameter}
    \end{table}
    
\subsubsection{Charge identification using muon capture in Argon}
The charge id of the muon can be identified using the capture vs decay process of the muon inside the argon\cite{PhysRevD.100.093004}. The working principle of charge id of the muon is as follows: a fraction of the $\mu^{-}$ like events that undergo the capture process are identified using capture fraction efficiency and the rest of the muons as well as all the $\mu^{+}$ undergo muon decay. We have implemented these as mentioned in \cite{Chatterjee:2022pqg}.\\
   
\subsection{$\chi^2$ analysis}
We have evaluated the $\chi^2$ for a fixed set of parameters using the method of pulls. We can tackle various statistical and systematic uncertainties directly through this method. The flux, cross sections, and other systematic uncertainties are taken into account by allowing these inputs to vary from their standard values in the computation of the expected rate in the $\text{i-j}^{\text{th}}$ bin, ${\mathrm{ N^{th}_{ij} }}$. Let the ${\mathrm{k^{th} }}$ input deviate from its standard value by ${\mathrm{\sigma_k \;\xi_k }}$ with ${\mathrm{ \sigma_k }}$ denoting the uncertainty. The value of ${\mathrm{ N^{th}_{ij} }}$ with the modified inputs is given by,
    
    \begin{equation}\label{eqn:cij}
    \mathrm{ N^{th}_{ij} =  N^{th}_{ij}(std) + \sum^{npull}_{k=1}
    c_{ij}^k \xi_k }\,,
    \end{equation}
    
where ${\mathrm{ N^{th}_{ij}(std) }}$ gives the expected rate  obtained with the standard values of the inputs in the $\text{i-j}^{\text{th}}$ bin and \textit{npull} is the total number of sources of uncertainty (5 for our case).
The ${\mathrm{ \xi_k }}$'s are the \emph{pull} variables and they determine the number of ${\mathrm{ \sigma}}$'s by which the ${\mathrm{ k^{th} }}$ input deviates from its standard value. In eq.~\eqref{eqn:cij}, ${\mathrm{ c_{ij}^k }}$ signifies the change in ${\mathrm{N^{th}_{ij} }}$ when the ${\mathrm{ k^{th} }}$ input is changed by ${\mathrm{ \sigma_k }}$ (i.e. by 1 standard deviation). The uncertainties in the inputs being not very large, we only consider changes in ${\mathrm{N^{th}_{ij} }}$ that are linear in ${\mathrm{ \xi_k }}$. Hence we evaluate the modified $\chi^2$ as,
    \begin{align}  \label{eqn:chisq}
        {\chi^2(\xi_k)} = &\sum_{i,j}\;
        \frac{\left[~N_{ij}^{th}(std) + \sum^{npull}_{k=1}
        c_{ij}^k \xi_k - N_{ij}^{ex}~\right]^2}{N_{ij}^{ex}}\nonumber\\&+\sum^{npull}_{k=1} \xi_k^2\,,
    \end{align}
    where the additional ${\rm{\xi_k^2}}$-dependent term is added due to the penalty imposed for moving the value of the ${\mathrm{k^{th}}}$ input away from its standard value by ${\rm{\sigma_k \;\xi_k}}$.  $\chi^2$ in the case of the standard LArTPC detector and for a detector with change id for muons are evaluated as,
    \begin{eqnarray}
        \chi^{2}_{standard} = \chi^{2}_{\mu^{-} + \mu^{+}} + \chi^{2}_{e^{-}+ e^{+}}\\
    \label{eq:chi_stnd}
        \chi^{2}_{charge-id} = \chi^{2}_{\mu^{-}}  + \chi^{2}_{ \mu^{+}} + \chi^{2}_{e^{-}+ e^{+}}
    \end{eqnarray}\label{eq:chi_chrg}
The $\chi^2$ with pulls, which includes the effects of all theoretical and systematic uncertainties, is obtained by minimizing ${\rm{\chi^2(\xi_k)}}$ with respect to all the pulls ${\rm{\xi_k}}$.
    \begin{equation}  \label{eqn:chisqmin}
        {\mathrm{ \chi^2_{pull} = Min_{\xi_k}\left[
        \chi^2(\xi_k)\right] }}\,.
    \end{equation}
Finally, We marginalize the $\chi^2$ over the allowed range of the oscillation parameters as mentioned in \autoref{table:chi-parameter}.
For the combined analysis we add the chi-square for beam and atmospheric and then marginalize over the oscillation parameters. The marginalization has been performed in $\theta_{23},\theta_{14},\theta_{24},\delta_{13},\delta_{14}$ over the range specified in \autoref{table:chi-parameter} for all cases unless otherwise mentioned. 
    \begin{table}
		\centering
		\begin{tabular}{|c|c|c|}
			\hline
			Parameter & True Values & Marginalization Range\\\hline
			$\theta_{12}$ & $33.47^\circ$ & N.A.\\
			$\theta_{13}$ & $8.54^\circ$ & N.A.\\
			$\theta_{23}$ & $45^\circ$ & $39^\circ:51^\circ$ \\$\theta_{14},\theta_{24}$ & $7^\circ$ & $0^\circ:10^\circ$\\
                $\theta_{14},\theta_{24}$ & $4^\circ$ & $0^\circ:6^\circ$\\
                $\theta_{34}$ & $0^\circ, 7^\circ$ & $0^\circ : 17^\circ$\\
			$\Delta_{21}$ & $7.42\times 10^{-5}$ eV$^2$ & N.A.\\$\Delta_{31}$(NO) & $2.5\times 10^{-3}$ eV$^2$ &$-(2.42:2.62)\times 10^{-3}$ eV$^2$\\$\Delta_{31}$(IO) & $-2.5\times 10^{-3}$ eV$^2$ &$(2.42:2.62)\times 10^{-3}$ eV$^2$\\
			$\Delta_{41}$ (for AMO) & $1$ eV$^2$ & N.A.\\
			$\Delta_{41}$ (for SMO) & $0.0005:0.01$ eV$^2$ & $\pm 15\%$ of $-\Delta_{41}$\\
			$\delta_{13}$ & many & $-180^\circ:180^\circ$\\ $\delta_{14}$ & $0^\circ,90^\circ,-90^\circ$ & $-180^\circ:180^\circ$\\
		    \hline
		\end{tabular}
		\caption{The table depicts true values of all the parameters and their range of marginalization as used in our analysis.}
		\label{table:chi-parameter}
    \end{table}
	
\section{Results and Discussion}
\label{subsec:res}

In this section, we present the results for the analysis of beam, atmospheric, and a combination of both data for the following cases,
\begin{itemize}
    \item determination of the sign of $\Delta_{31}$ (AMO) for $\Delta_{41}=1$ eV$^2$

    \item determination of the sign of $\Delta_{31}$ for $\Delta_{41}$ in the range of $5\times10^{-4}: 0.1$ eV$^2$

    \item determination the sign of $\Delta_{41}$ (SMO) when it's value lies in the range of $5\times10^{-4}: 0.1$ eV$^2$
\end{itemize}

\subsection{Sensitivity to the AMO for $|\Delta_{41}|=1$ eV$^2$}
    \begin{figure}
		\centering
		\includegraphics[width=0.96\linewidth]{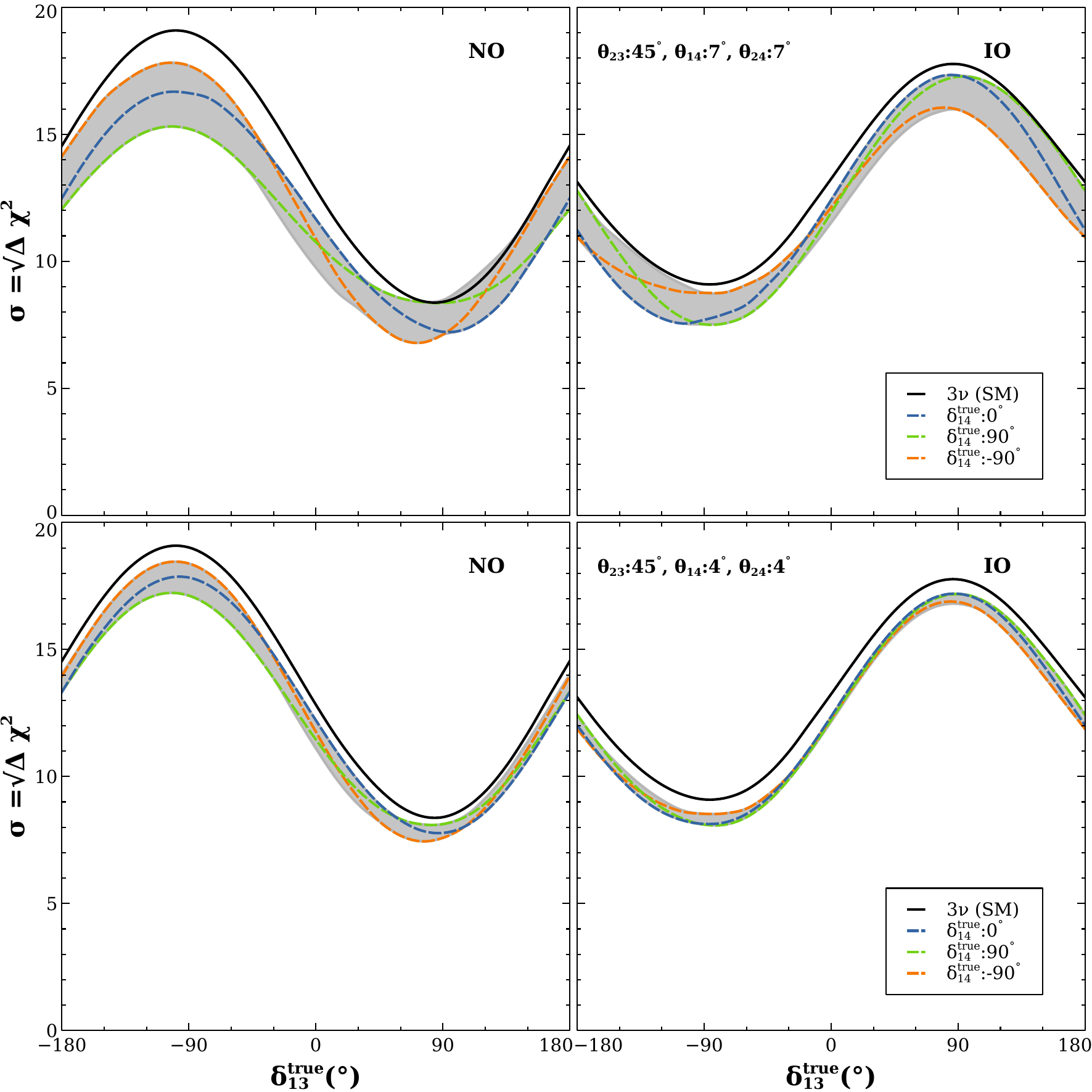}
		\caption{The sensitivity to the atmospheric mass ordering as a function of true $\delta_{13}$ for various $\delta_{14}^{true}$ values at 1300 km baseline considering normal (left), inverted (right) ordering.}
		\label{fig:sigma_d13-d14_var}
    \end{figure}
In fig.~\ref{fig:sigma_d13-d14_var}, the sensitivity to the atmospheric mass ordering (AMO) is presented as a function of $\delta_{13}^{true}$ in standard three flavor framework (black) for normal (left) and inverted (right) ordering. We also present the sensitivity in the presence of a sterile neutrino corresponding to SNO-NO (left), and SNO-IO (right) for true values of $\delta_{14}=0^\circ$(blue), $90^\circ$(green), $-90^\circ$(orange), $180^\circ$(red). We will call $-180^\circ<\delta_{13}\leq 0^\circ$ as the lower half plane[LHP] and $0^\circ<\delta_{13}\leq 180^\circ$ as the upper half plane[UHP] throughout this section. The important points to be noted are,
    \begin{itemize}
        \item The sensitivity decreases in the presence of a sterile neutrino compared to the three flavor case.
        
        \item The sensitivity for the sterile cases depends on the true values of $\delta_{14}, \delta_{13}$.

        \item For $\theta_{14},\theta_{24}=4^\circ$, the sensitivity is higher than that of $\theta_{14},\theta_{24}=7^\circ$ and also closer to the standard 3$\nu$ case. This is due to the fact that the smaller the sterile mixing angles are the more the sensitivity of the sterile case gets closer to the results of the standard $3\nu$ case.  
        
        \item For NO, in the LHP of true $\delta_{13}$ the highest sensitivity is observed for $\delta_{14}^{true}=-90^\circ$, and the lowest sensitivity for $\delta_{14}^{true}=90^\circ$ whereas in the UHP, this order gets reversed.
        
        \item For IO, $\delta_{14}^{true}=0^\circ$ (blue) shows the lowest sensitivity in the LHP of true $\delta_{13}$ and also the highest sensitivity in UHP.
    \end{itemize}

\begin{figure}[H]
        \centering
        \includegraphics[width=0.96\linewidth]{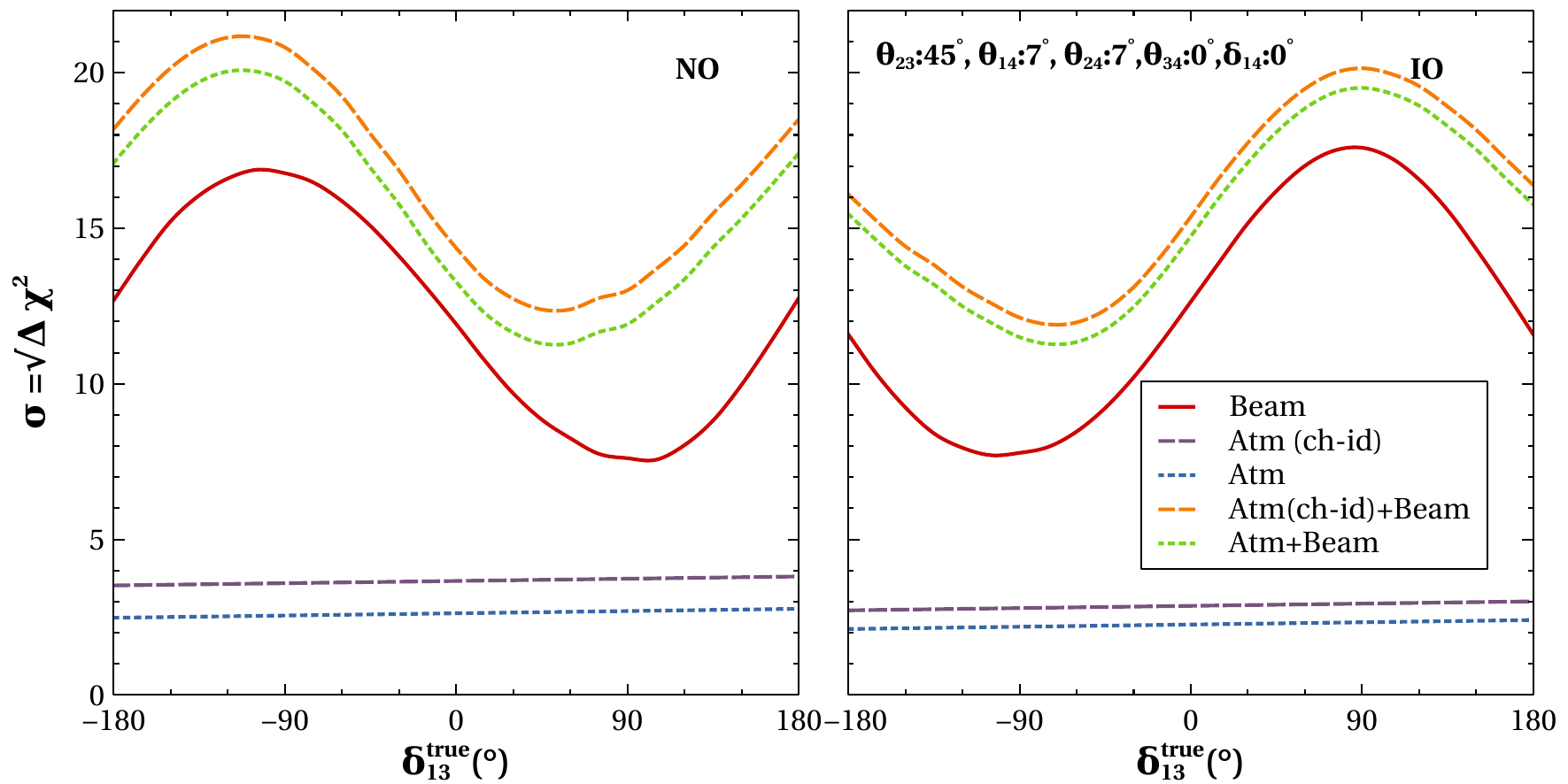}
        \caption{Atmospheric mass ordering sensitivity as a function of $\delta_{13}^{true}$ corresponding to the analysis of  only beam (red), only atmospheric (violet), combined atmospheric+beam (green) neutrinos for normal (left) and inverted (right) hierarchy with 400 kt-yr exposure of LArTPC}
        \label{fig:MH_Atmos-charge-id}
    \end{figure}

Next, in fig.~\ref{fig:MH_Atmos-charge-id}, the AMO sensitivity is shown as a function of true $\delta_{13}$ corresponding to the analysis of only beam(red), only atmospheric (blue) and a combination of both beam and atmospheric(green) simulated data. The cases with charge identification in atmospheric only (violet) and combined analysis (orange) are also depicted. The representative sensitivity curves are obtained for $\theta_{14},\theta_{24}=7^\circ$, $\delta_{14}=0^\circ$ corresponding to true hierarchy considered as normal (left) and inverted (right). The observations from fig.~\ref{fig:MH_Atmos-charge-id} are following,
    \begin{itemize}
        \item Sensitivity for atmospheric neutrinos doesn't have significant dependence on $\delta_{13}$.
        
        \item Combined sensitivity of beam and atmospheric is greater than the sum of individual sensitivities of these two, which demonstrates the synergy between them.

        \item We observe slightly higher sensitivity when we use partial charge identification for atmospheric neutrinos. This also leads to higher sensitivity for combined analysis.
    \end{itemize}

In fig.~\ref{fig:sigma_d13-th14,24,34_var}, we present the effect of $\theta_{34}$ on the sensitivity to the atmospheric mass ordering. In this plot, the sensitivity is shown as a function of true $\delta_{13}$ for various combinations of true values of $\theta_{34},\delta_{34}$. We consider for beam only analysis, $\theta_{34}=0^\circ$, $\delta_{34}=0^\circ$(green dotted), and $\theta_{34}=7^\circ$,$\delta_{34}=90^\circ$(red dotted) along with the sensitivity curve for standard three flavors (black). We also plot the sensitivity with combined beam and atmospheric analysis $\theta_{34}=7^\circ$,$\delta_{34}=90^\circ$ (red-dashed). Other sterile parameters are fixed as $\delta_{14}=90^\circ$, $\theta_{14}=\theta_{24}=7^\circ$. The observations are as follows,
    \begin{itemize}
        \item The sensitivity decreases as $\theta_{34}$ becomes higher.
        
        \item In the combined analysis of beam and atmospheric data, the sensitivity is higher than the beam analysis compensating for the decrease due to non-zero $\theta_{34}$.
    \end{itemize}
    
    \begin{figure}[H]
        \centering
        \includegraphics[width=0.96\linewidth]{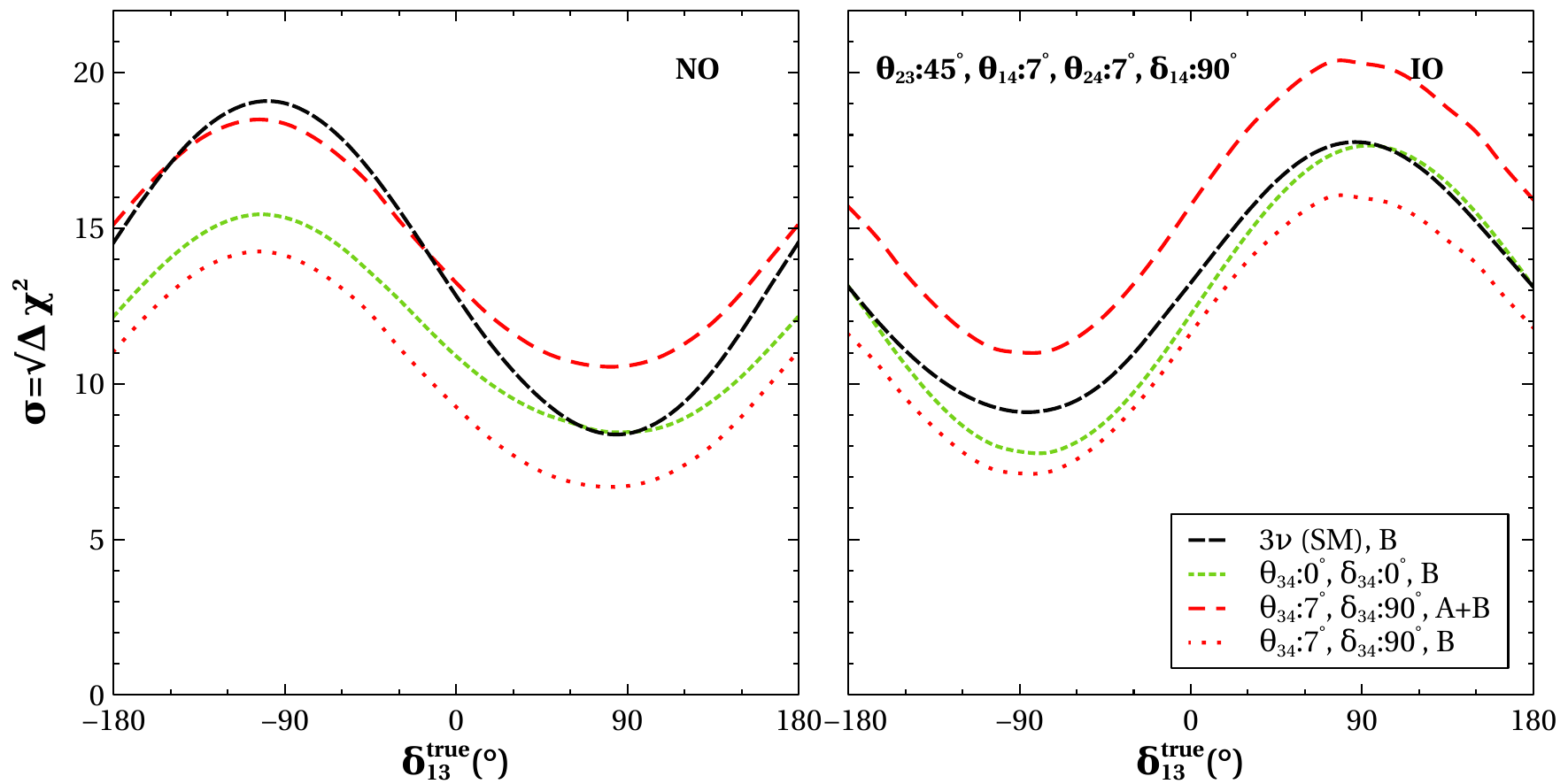}
        \caption{The sensitivity to mass ordering as a function of $\delta_{13}^{true}$ for various $\theta_{34}^{true}$ values for beam neutrinos with 1300 km baseline for normal (left), inverted (right) ordering}
        \label{fig:sigma_d13-th14,24,34_var}
    \end{figure}

\subsection{Sensitivity of the sign of $\Delta_{31}$ for $\Delta_{41}=10^{-4}:10^{-1}$ eV$^2$ }

In this section, we study how the sensitivity to the sign of $\Delta_{31}$ behaves with $\Delta_{41}$ where the latter varies in the range of $10^{-4}:10^{-1}$ eV$^2$. Note that for $\Delta_{41} \sim 1 $ eV$^2$ only SNO-NO, SNO-IO cases are cosmologically allowed. However, for $\Delta_{41}=10^{-4}:10^{-1}$ eV$^2$ all the four possibilities depicted in fig.~\ref{fig:ster_mass} are admissible.
    \begin{figure}[H]
        \centering
        \includegraphics[width=0.96\linewidth]{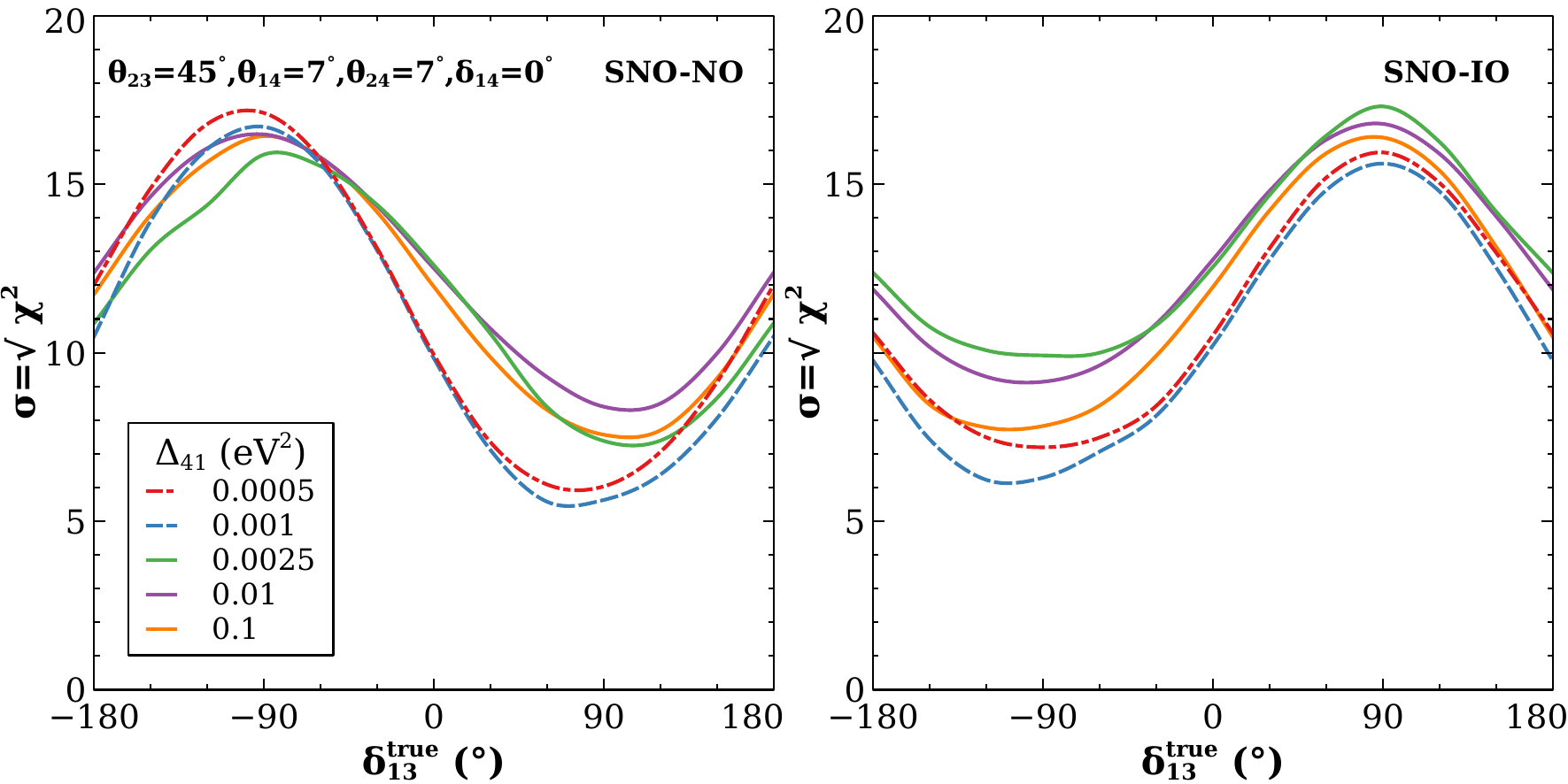}
        \includegraphics[width=0.96\linewidth]{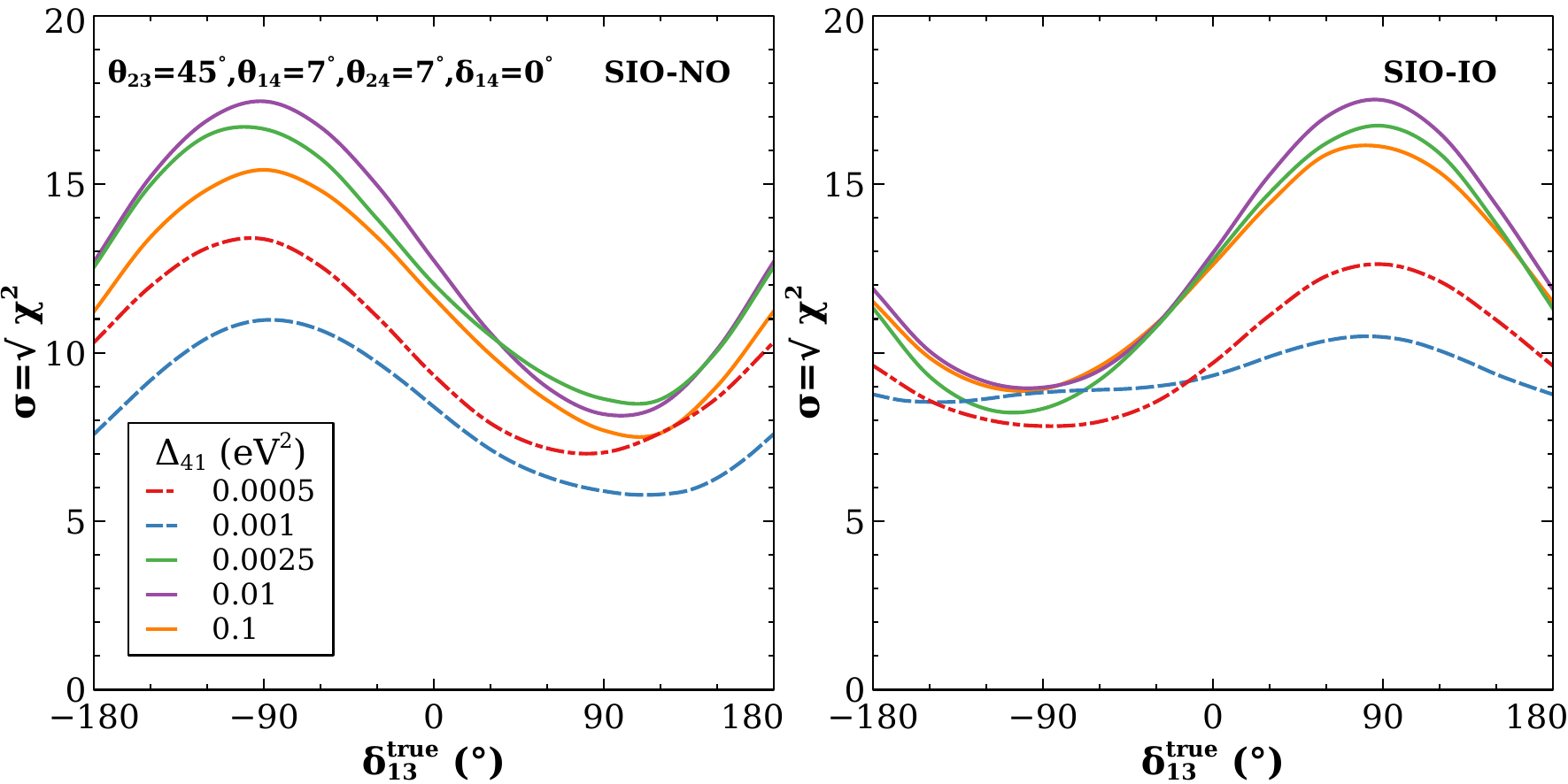}
        \caption{Sensitivity to atmospheric mass ordering as a function of $\delta_{13}^{true}$ in SNO(top), SIO (bottom) scenarios with true $\Delta_{31}$  for different values of $\Delta_{41}^{true}$ at 1300 km baseline}
        \label{fig:sig-mh-d13_D41-D}
    \end{figure}
In fig.~\ref{fig:sig-mh-d13_D41-D}, the AMO sensitivity is shown as a function of $\delta_{13}^{true}$ at various true values of $\Delta_{41}$. The upper(lower) panels correspond to the true value in SNO (SIO) cases while the left (right) panels are for NO(IO). During the computation of $\chi^2$, the $|\Delta_{41}|$ is fixed in true and test cases for this plot. The observations of significance in fig.~\ref{fig:sig-mh-d13_D41-D} are as follows,
\begin{itemize}
        \item The nature of variation of sensitivity with $\delta_{13}^{true}$ doesn't change significantly for different true values of $\Delta_{41}$. 

        \item Sensitivity gets notably reduced for $\Delta_{41}=0.001$ eV$^2$(blue) for the most of values of $\delta_{13}^{true}$ in the UHP in SNO-NO and SIO-IO case. In SNO-IO and SIO-NO cases, blue curves give minimum sensitivity over the full range of $\delta_{13}$

        \item However, sensitivity for $\Delta_{41}=0.001$ eV$^2$ is very high in the LHP of $\delta_{13}^{true}$ in SNO-NO, SIO-IO.
        
        \item The maximum sensitivity is observed for $\Delta_{41}=0.01$ eV$^2$(violet) in SIO-NO and SIO-IO case over almost the full range of $\delta_{13}^{true}$.

        %\item For SIO-IO case $\Delta_{41}=2.5\times 10^{-3}$ eV$^2$ shows the maximum sensitivity over full range of $\delta_{13}^{true}$.
    \end{itemize}

\begin{figure}[H]
    \centering
    \includegraphics[width=0.96\linewidth]{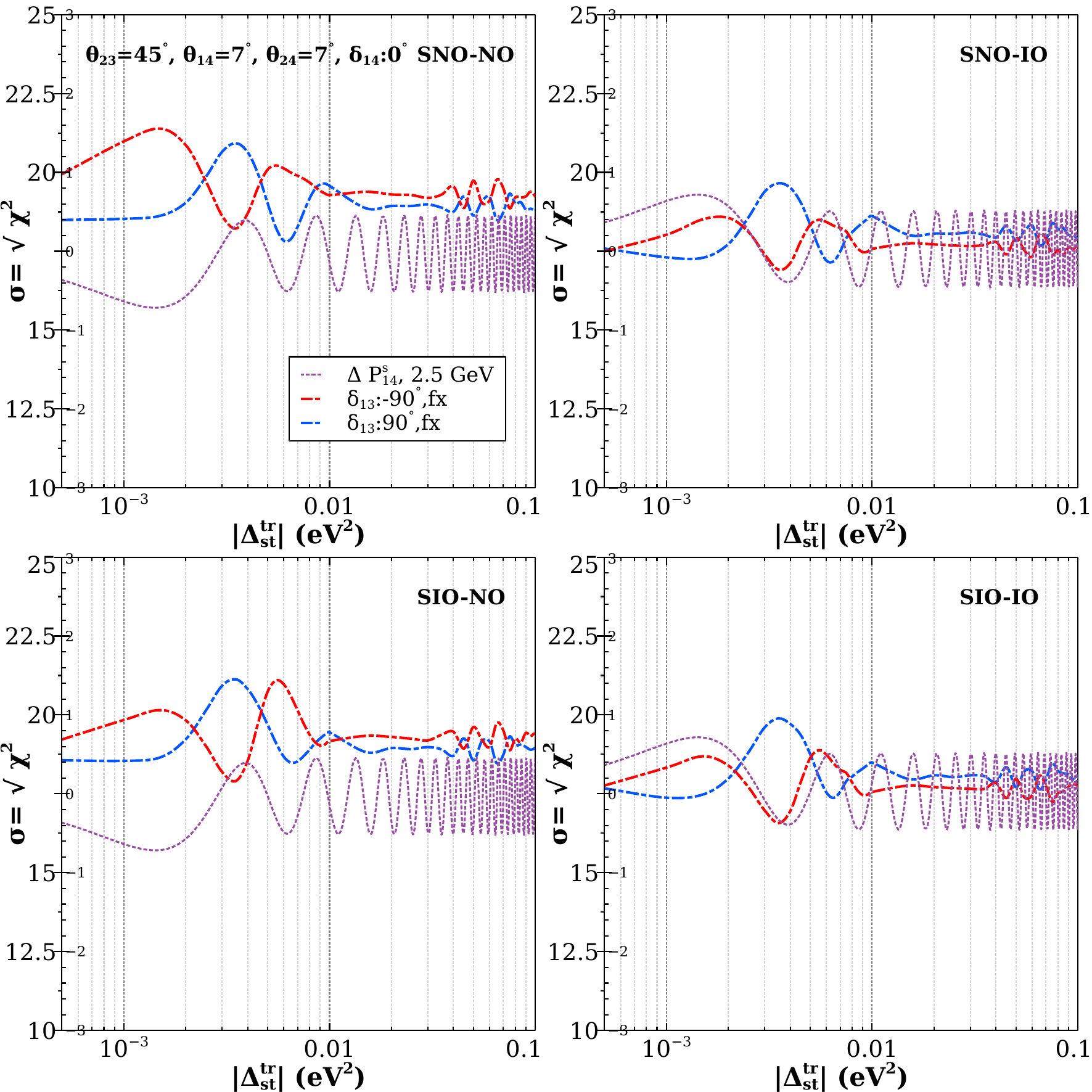}
    \caption{Sensitivity to atmospheric mass ordering as a function of $\Delta_{41}^{true}$ with marginalisation in $\Delta_{31}$ for $\theta_{23}^{true}=45^\circ$, $\delta_{14}^{true}=0^\circ$, $\delta_{13}^{true}=-90^\circ$(red), $90^\circ$(blue) at 1300 km baseline. The violet curve shows $\Delta P_{14}^s$ at 2.5 GeV.}
    \label{fig:sig-fx_dpme14s-ana}
\end{figure}
We note that for the difference in the probability between the opposite AMO, the dependence on $\Delta_{41}$ will come from the last term in eq. \eqref{eq:Pme_alpha}. The probability difference for fixed values of $\theta_{13},\theta_{23},\theta_{14},\theta_{24},\delta_{13},\delta_{14}$ and a given SMO can be expressed as, 
    \begin{eqnarray}\label{eq:dPme_st}
        \Delta P_{\mu e}^{st} &= P_{\mu e}^{m,true}(i_1 \Delta_{41}, j_1 \Delta_{31}) - P_{\mu e}^{m,test}(i_2 \Delta_{41}, j_2 \Delta_{31})\nonumber\\
        &\propto 4 s_{13} s_{14} s_{24} s_{23}[\Delta P_{14}^s \sin{\tilde{\delta}_{14}} + \Delta P_{14}^c \cos{\tilde{\delta}_{14}]} \text{,}
    \end{eqnarray}
where we define;
    \begin{eqnarray}\label{eq:dpme14-c,s}
        \Delta P_{14}^{s,c} &= \frac{\sin[(A'-1)\Delta]}{(A'-1)} P_{14}^{s,c}(i_1 \Delta_{41}, j_1 \Delta_{31})\nonumber\\&- \frac{\sin[(A'+1)\Delta]}{-(A'+1)} P_{14}^{s,c}(i_2 \Delta_{41}, j_2 \Delta_{31})
    \end{eqnarray}
Depending upon the true value considered in a certain mass ordering, there can be four scenarios as follows
\begin{itemize}
    \item SNO-NO: $i_1=i_2=$+ve, $j_1=$+ve and $j_2=$-ve
    \item SNO-IO: $i_1=i_2=$+ve, $j_1=$-ve and $j_2=$+ve
    \item SIO-NO: $i_1=i_2=$-ve, $j_1=$+ve and $j_2=$-ve
    \item SIO-IO: $i_1=i_2=$+ve, $j_1=$+ve and $j_2=$-ve
\end{itemize}
Note that $\Delta P_{14}^{s,c}$ for different cases are connected as,
\begin{eqnarray}
    \Delta P_{14}^{s,c}|_{\rm SNO-NO}&=-\Delta P_{14}^{s,c}|_{\rm SNO-IO}\label{eq:dpme14-c,s-sno},\\
    \Delta P_{14}^{s,c}|_{\rm SIO-NO}&=-\Delta P_{14}^{s,c}|_{\rm SIO-IO}\label{eq:dpme14-c,s-sio}
\end{eqnarray}

In fig.~\ref{fig:sig-fx_dpme14s-ana}, we have depicted the sensitivity to AMO as a function of $|\Delta_{41}|$ with marginalization performed only over $\Delta_{31}$ with all other parameters being fixed. The red (blue) curve refers to $\delta_{13}=-90^\circ(90^\circ)$. We also show the difference in probability term $\Delta P_{14}^{s}$\eqref{eq:dpme14-c,s} evaluated at 2.5 GeV by the violet curve. The understandings from fig.~\ref{fig:sig-fx_dpme14s-ana} are as follows,
\begin{itemize}
    \item Since we have chosen $\delta_{14}=0^\circ$ for $\delta_{13}=90^\circ$ and $-90^\circ$, the value of $\sin\tilde{\delta}_{14}=\sin[\delta_{13}+\delta_{14}]$ is +1, -1 respectively and $\cos\tilde{\delta}_{14}=0$. For the fixed phases and mixing angles in both true and test cases, the difference in probability\eqref{eq:dPme_st} between NO and IO will only depend on $\Delta P_{14}^s$ as,
    \begin{equation}\label{eq:dPme_st_fin}
        \Delta P_{\mu e}^{st}= 4 s_{13} s_{14} s_{24} s_{23}\Delta P_{14}^s \sin\tilde{\delta}_{14}
    \end{equation}
   This means $\Delta P_{\mu e}^{st}$ will be opposite for $\delta_{13}=90^\circ$ and $-90^\circ$ leading to the opposite nature of chi-square.
    
    \item In the case of SNO-NO, for $\delta_{13}=90^\circ$, we have $\Delta P_{\mu e}^{st}\propto\Delta P_{14}^s|_{\rm SNO-NO}$. This can be seen from the top-left panel which shows that the nature of the blue and violet curves are similar. For $\delta_{13}=-90^\circ$, $\Delta P_{\mu e}^{st}\propto -\Delta P_{14}^s|_{\rm SNO-NO}$ which is reflected in the red curve being opposite to the violet curve.

    \item The sensitivity for SNO-IO is just opposite in nature to SNO-NO as shown in eq.~\eqref{eq:dpme14-c,s-sno}. Therefore, in the top-right panel, the red curve for $\delta_{13}=-90^\circ$ is similar to the violet curve here, and the blue curve for $\delta_{13}=+90^\circ$ is the opposite of the violet curve.
    
    \item In SIO-NO, and SIO-IO cases, we also observe similar patterns as SNO-NO and SNO-IO, respectively.

    \item The sensitivity is seemed to be almost constant for $R>>1$, i.e., $\Delta_{41}>>\Delta_{31}$. This can be understood analytically as follows. The difference in $P_{\mu e}$ for fixed energy and phases will only depend on $\Delta P_{14}^{s}$ and can be evaluated using eq. \eqref{eq:dpme14-c,s} for $\Delta_{41}>>1$ limit as, 
    \begin{eqnarray}\label{eq:dP14s_r>>1}
        \Delta P_{14}^{s}&= \left(\frac{\sin[(A'-1)\Delta]}{(A'-1)} + \frac{\sin[(A'+1)\Delta]}{(A'+1)}\right)\nonumber\\&\times\Big\{(\frac{A'}{2R}c_{23}+2) \sin[R\Delta]^2 \Big\},
    \end{eqnarray}
    where the term in the braces is dependent on $\Delta_{41}$. For $R>>1$ the term $\sin[R\Delta]$ shows fast oscillation. Summing over energies, $\sin[R\Delta]$, will give a constant value that is reflected in the sensitivity curves.

\end{itemize}

\begin{figure}[H]
            \centering
            \includegraphics[width=0.96\linewidth]{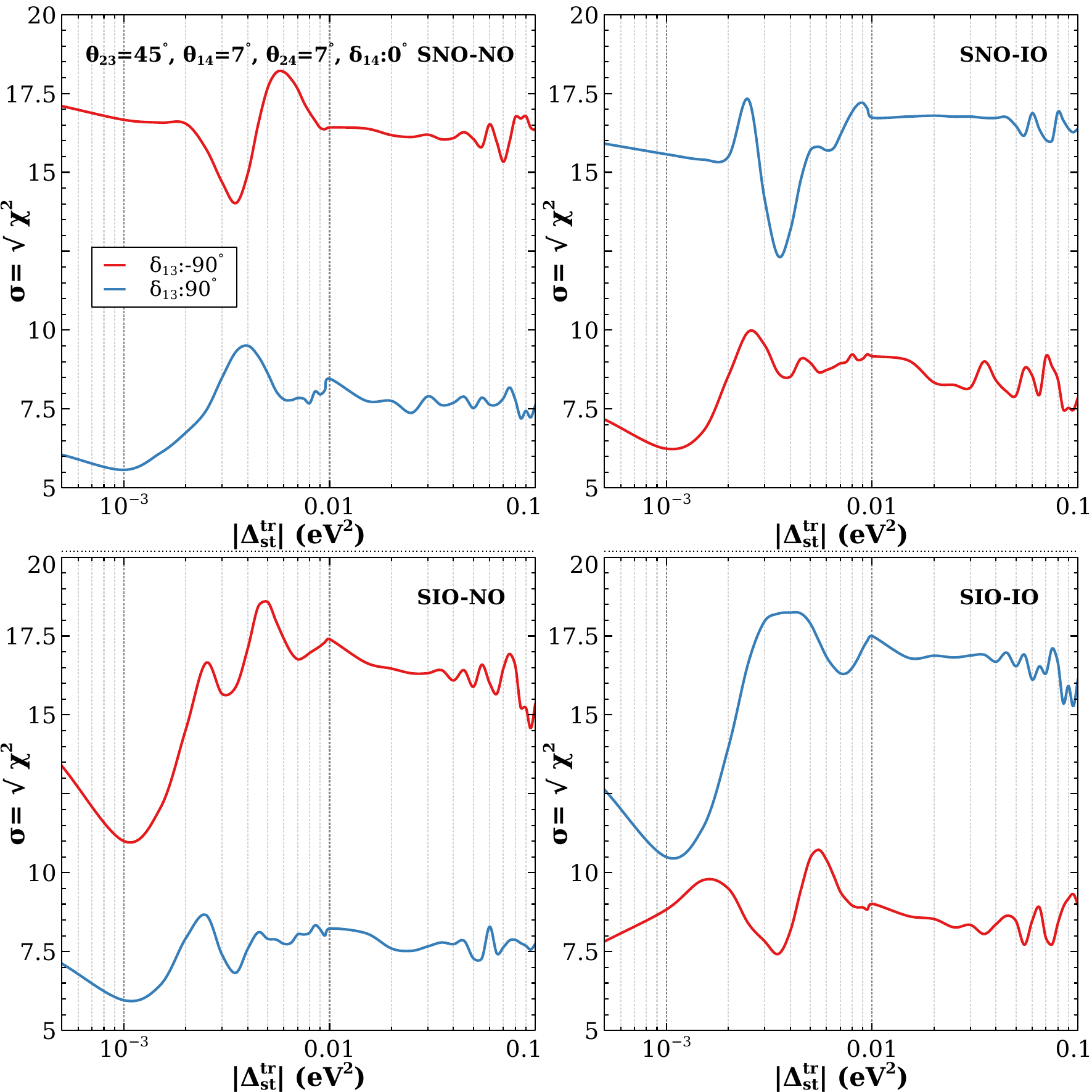}
            \caption{Sensitivity to atmospheric mass ordering as a function of $\Delta_{41}^{true}$ for $\theta_{23}^{true}=45^\circ$, $\delta_{14}^{true}=0^\circ$, $\delta_{13}^{true}=-90^\circ$ (red), $90^\circ$(blue) at 1300 km baseline}
            \label{fig:sig-D41-mo-beam}
    \end{figure}

In fig.~\ref{fig:sig-D41-mo-beam}, we depict the sensitivity to the sign of $\Delta_{31}$ as a function of true $\Delta_{41}$ for $\delta_{13}=-90^\circ(\rm{red}),90^\circ(\rm{blue})$, and $\delta_{14}=0^\circ$ at 1300 km. For this figure, we have marginalized over the parameters as mentioned in table`~\ref{table:chi-parameter}.
Some interesting features of sensitivity to the AMO as seen from fig.~\ref{fig:sig-D41-mo-beam} are as follows,
\begin{itemize}
    \item For SNO-NO (top-left panel) and SIO-IO (bottom-right panel) cases we observe a contrasting nature of the sensitivity between $\delta_{13}=90^\circ$ and $-90^\circ$. Note that a similar contrasting nature has been observed in the top two panels of fig. \ref{fig:dPme-mo-beam} showing the oscillogram of $\Delta P_{\mu e}$. For instance, in SNO-NO, at $\Delta_{41}=3\times10^{-3}$ eV$^2$ a maxima of sensitivity occurs for $\delta_{13}=90^\circ$ whereas minima occurs for $\delta_{13}=-90^\circ$.
    
    \item However, the nature of the sensitivity curves for $\delta_{13}=90^\circ$ and $-90^\circ$ is similar in SNO-IO (top-right panel) and SIO-NO (bottom-left panel) cases. This behavior was also observed in  $\Delta P_{\mu e}$ oscillogram in the bottom panels of fig. \ref{fig:dPme-mo-beam} for SNO-IO case.
    
    \item In all the cases the minima and maxima are observed in the range of $0.001-0.01$ eV$^2$. Beyond that, the sensitivity is relatively flat with $\Delta_{41}$.
\end{itemize}

\subsection{Sensitivity to the sign of $\Delta_{41}$ (SMO)}
\begin{figure}
        \centering
        \includegraphics[width=0.96\linewidth]{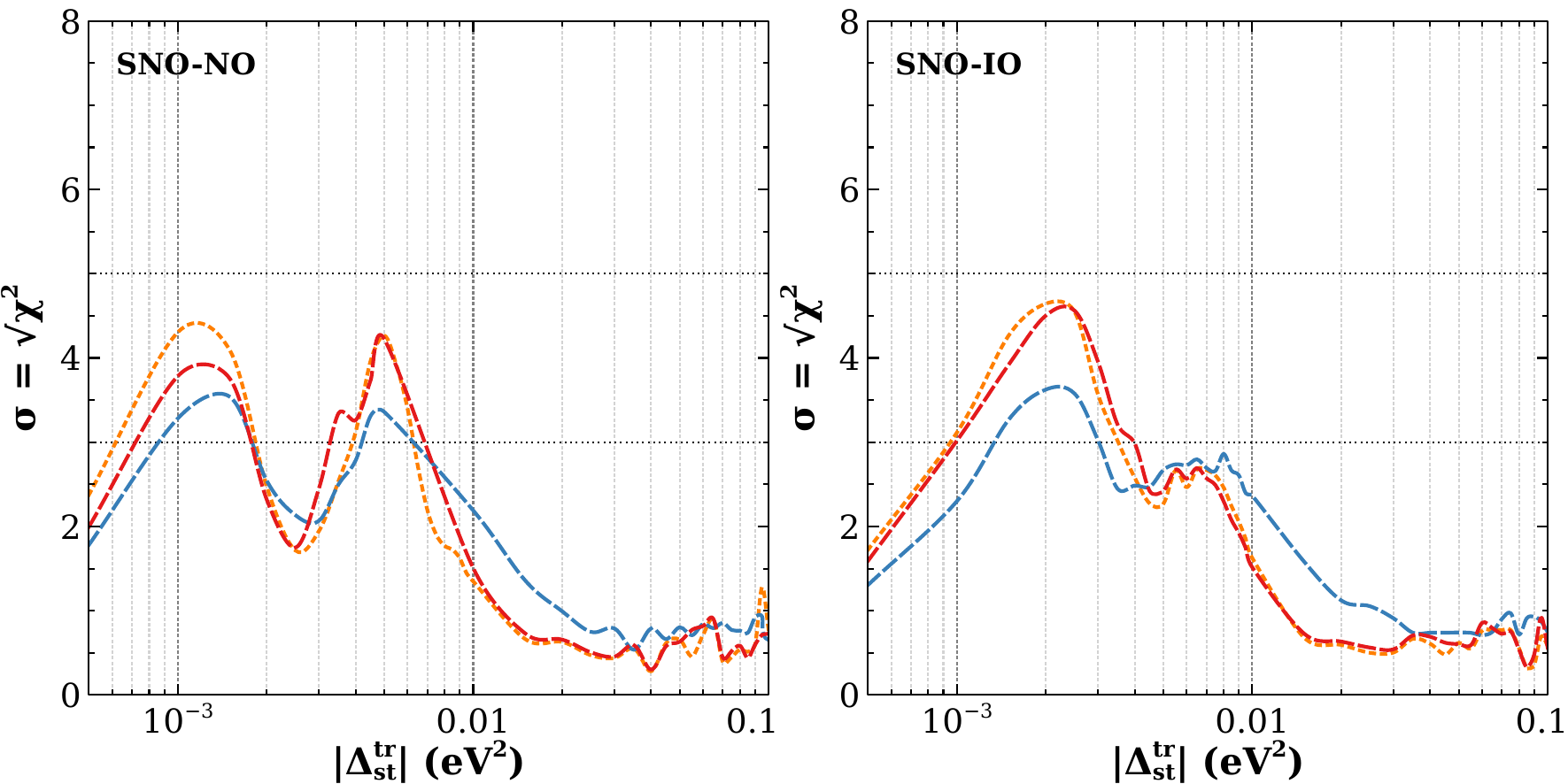}
        \includegraphics[width=0.96\linewidth]{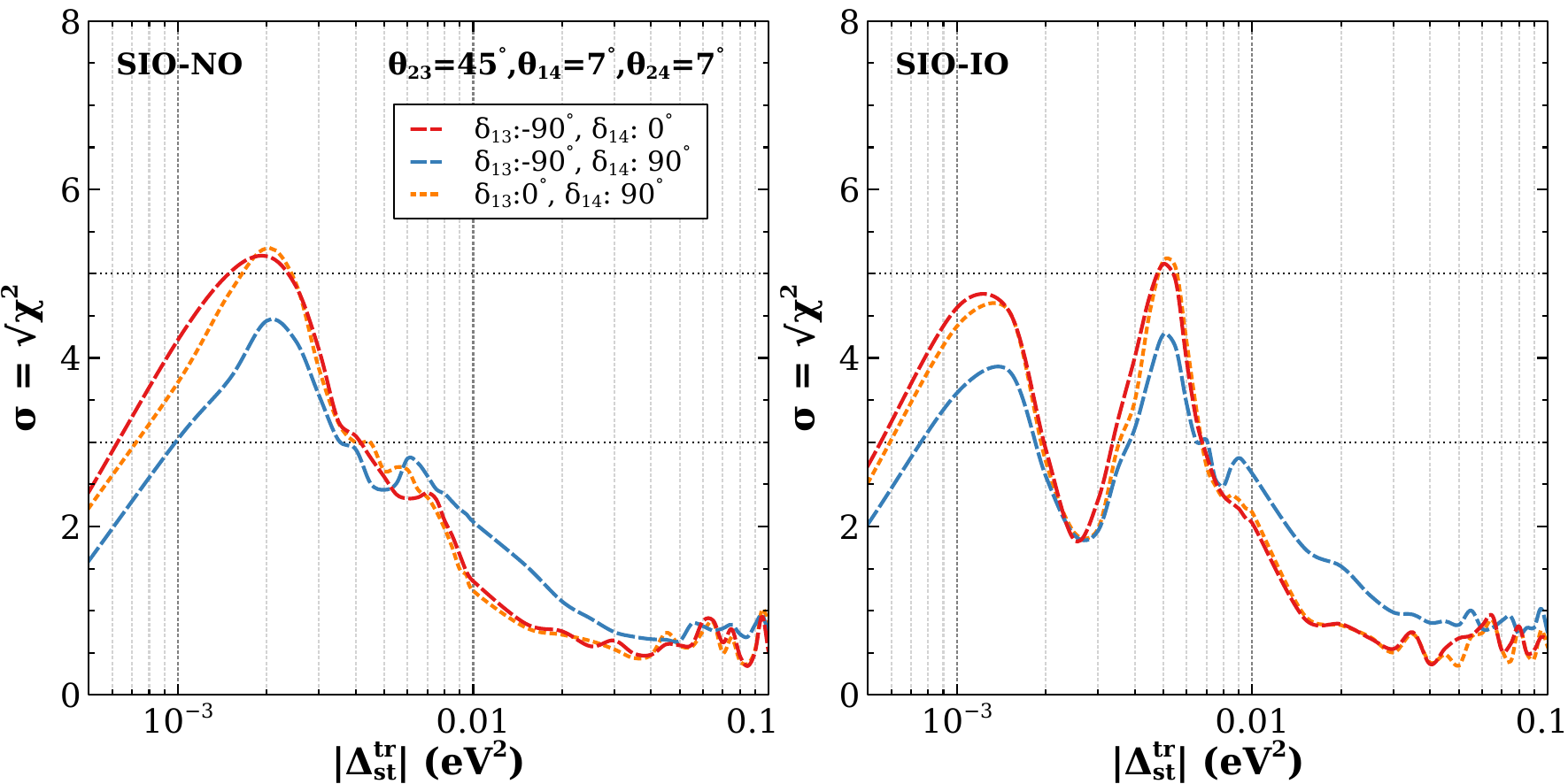}
        \caption{Sensitivity to sterile mass ordering as a function of $\Delta_{41}^{true}$ for $\Delta_{31}=+ve$(left), $-ve$(right) and  $\Delta_{41}=+ve$(top), $-ve$(bottom) using different values of $\delta_{14}^{true}$, $\delta_{13}^{true}$ at 1300 km.}
        \label{fig:st-mh-hr-beam}
    \end{figure}

In this section, we present the sensitivity  of the sign of $\Delta_{41}$ considering it's values in the range of $[5\times10^{-4}:10^{-1}]$ eV$^2$ for which all four possibilities depicted in fig 1 will be viable. In fig.~\ref{fig:st-mh-hr-beam}, the sensitivity of the sign of sterile mass squared differences are depicted as a function of the true value of $\Delta_{41}$ for various true values of $\delta_{13},\delta_{14}$\footnote{This plot has been presented in ref \cite{Chattopadhyay:2022hkw} for fixed values of $\delta_{13}=0^\circ,\delta_{14}=0^\circ$}.
The salient features of the fig.~\ref{fig:st-mh-hr-beam} are as follows,
    \begin{itemize}
        \item The sensitivity curves for SNO-NO and SIO-IO show two prominent maxima around true values of $\Delta_{41}=1\times 10^{-3}$ eV$^2$, $5\times 10^{-3}$ eV$^2$ cases. There is a dip in sensitivity when true $\Delta_{41}$ is around $2.5\times10^{-3}$ eV$^2$ due to its proximity to atmospheric mass squared difference as discussed in ref.~\cite{Chattopadhyay:2022hkw}.
        
        \item In the case of SNO-IO and SIO-NO, the maxima occurs around $\Delta_{41}=2.5\times 10^{-3}$ eV$^2$, i.e, when the sterile mass squared difference is close to the atmospheric mass squared difference $\Delta_{31}$.

        \item The SNO-IO and SIO-IO cases provide relatively higher sensitivity than the SNO-NO and SIO-IO cases. 

        %\item We also observe a variation of $1\sigma$ in sensitivity at a fixed $\Delta_{41}$ for different values of the phases $\delta_{13}, \delta_{14}$.
        
        \item The features of sensitivity in SNO-NO can be qualitatively understood from the plot of probability difference in top panels of fig.~\ref{fig:dPme-st-mh-hr-beam}.
    \end{itemize}
    \begin{figure}
        \centering
        \includegraphics[width=0.96\linewidth]{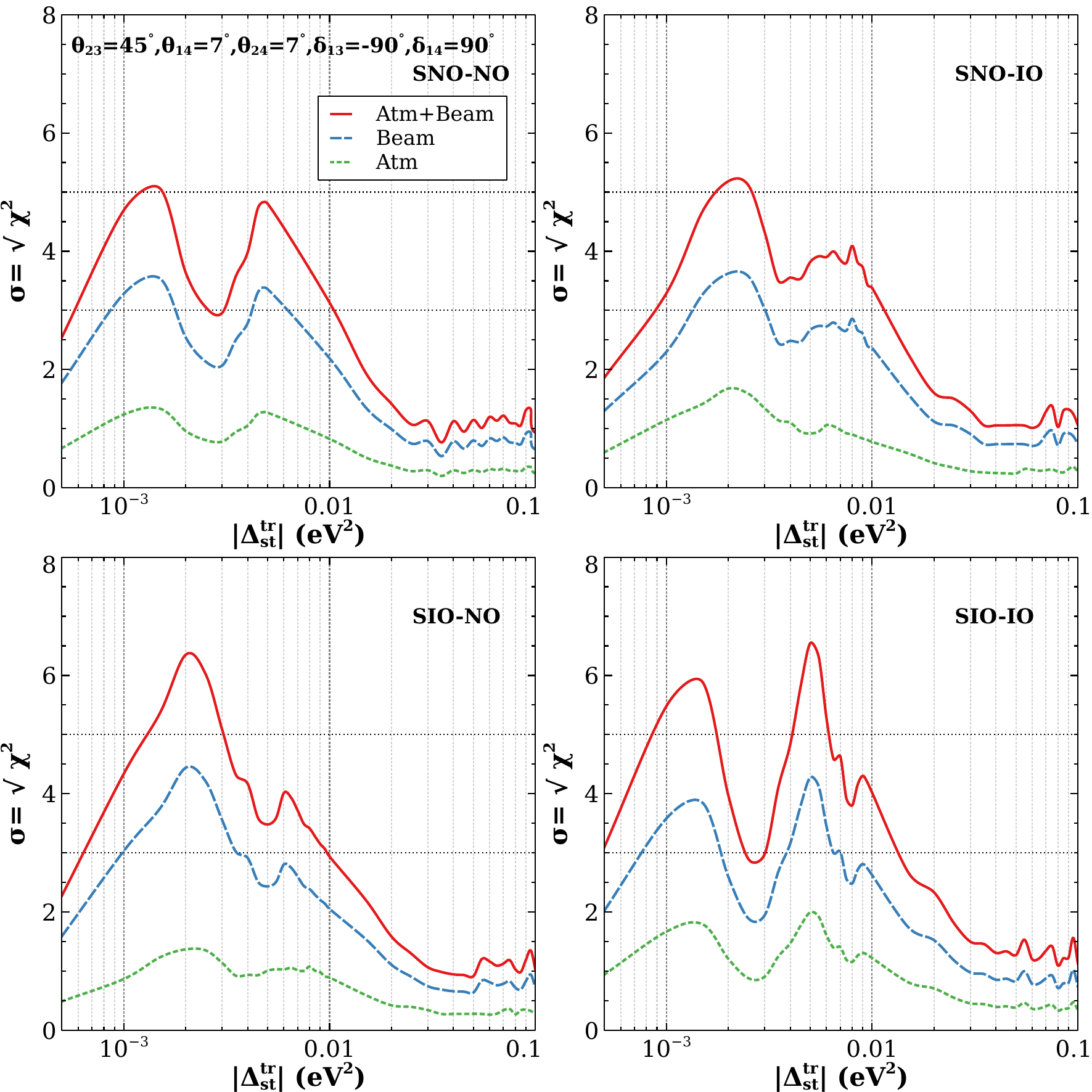}
        \caption{Sensitivity to sterile mass hierarchy as a function of $\Delta_{41}^{true}$ using combined beam and atmospheric neutrinos at 1300 km baseline.}
        \label{fig:st-mh-hr-atm}
    \end{figure}
In fig.~\ref{fig:st-mh-hr-atm}, we have used the simulated data from atmospheric neutrino analysis to perform a combined analysis of beam and atmospheric neutrinos and get the sensitivity of the sterile mass ordering as a function of the true value of $\Delta_{41}$ for the true value of phases $\delta_{13}=-90^\circ$(left), $\delta_{14}=90^\circ$(right). In these four panels, it is observed that the sensitivity to SMO is bettered in combined analysis than the beam neutrinos. The nature of the sensitivity is almost similar for beam and atmospheric analysis. This can be understood from the similar profile of difference in probabilities $\Delta P_{\nu e}, \Delta P_{\mu \mu}$ at 1300 km and 7000 km as shown in fig.~\ref{fig:dPme-mo-beam}. The combined sensitivity is above $3\sigma$ for most of the parameter space up to $\Delta_{41}=10^{-2}$ eV$^2$. For the values of $\Delta_{41}$ greater than that, even with the addition of atmospheric neutrinos, we get fixed sensitivity of $1.5\sigma-2\sigma$.
    
\section{Conclusions}
\label{subsec:con}
Our work focuses on the effect of an additional light sterile neutrino with a mass squared difference in the range of $10^{-4}:1$ eV$^2$ on the determination of atmospheric mass ordering and sterile mass ordering. For our study, we use a liquid argon detector for a) beam neutrinos (baseline 1300 km), and b) atmospheric neutrinos. 

The main new aspect of our study is to do a combined analysis of beam and atmospheric neutrinos to determine the sign of $\Delta_{31}$ and $\Delta_{41}$ when the latter is varied in the range $10^{-4}-1 $ eV$^2$. We also study in detail the impact of various sterile parameters.

The sensitivity to the sign of $\Delta_{31}$(AMO) for $\Delta_{41}=1$ eV$^2$ gets diminished w.r.t. to the 3$\nu$ case in the presence of a sterile neutrino. The decrement in the sensitivity is higher for larger values of $\theta_{14},\theta_{24}$. The values of $\chi^2$ also depend on $\delta_{13},\delta_{14}$. The sensitivity to AMO decreases further in the presence of a non-zero $\theta_{34}$. However, with the combined analysis of beam and atmospheric neutrino, we are able to recover the sensitivity over 10$\sigma$, irrespective of the choice of true values of $\delta_{13},\delta_{14}$. The presence of a light sterile neutrino gives the possibility of both positive and negative values of $\Delta_{41}$. Our study also demonstrates for the first time the dependence of the atmospheric mass ordering sensitivity on the absolute value of $\Delta_{41}$ as well as the on the nature of the 3+1 mass spectrum (SNO-NO, SNO-IO, SIO-NO, SIO-IO).

We also study the sensitivity to the sign of $\Delta_{41}$ (SMO) for the $\Delta_{41}=5\times10^{-4}:0.1$ eV$^2$ in different scenarios of the 3+1 mass spectrum. The sensitivity gets reduced when $\Delta_{41}$ is in the proximity of $\Delta_{31}$ for the SNO-NO and SIO-IO cases whereas in the SNO-IO and SIO-NO cases, the sensitivity gets enhanced. The addition of atmospheric neutrinos boosts the sensitivity over 3$\sigma$ for $\Delta_{41}<10^{-2}$ eV$^2$. However, for higher values of $\Delta_{41}$, the sensitivity falls off to $\sim 1.5\sigma-2\sigma$ for the combined analysis.

\acknowledgments
AC acknowledges the University of Pittsburgh, where the initial phase of the work has been done. AC also acknowledges the Ramanujan Fellowship (RJF/2021/000157), of the Science and Engineering Research Board of the Department of Science and Technology, Government of India. AC also acknowledges CERN for providing facilities. SG acknowledges the J.C. Bose Fellowship (JCB/2020/000011) of the Science and Engineering Research Board of the Department of Science and Technology, Government of India. It is to be noted that this work has been done solely by the authors and is not representative of the DUNE collaboration.

% query inspire, trying to verify the data (this will probably spot
% eventual typos) and retrieve the document DOI and eventual errata.
% in short all the information that clearly identify a document.

\bibliographystyle{apsrev4-1}
\bibliography{references}
%\begin{thebibliography}{99}
%\end{thebibliography}

\end{document}